\documentclass[aps,prd,12pt,nofootinbib]{revtex4}

\usepackage{graphicx}
\usepackage{amsmath}
\usepackage{array}
\usepackage{color}

\usepackage{soul,xcolor}
\setstcolor{red}

\newcommand{\be}{\begin{equation}}
\newcommand{\ee}{\end{equation}}
\newcommand{\ba}{\begin{eqnarray}}
\newcommand{\ea}{\end{eqnarray}}
\newcommand \nn {\nonumber}

\begin{document}

\title{
Kubo formulas for the shear and bulk viscosity relaxation times
and the scalar field theory shear $\tau_\pi$ calculation}

\author{Alina Czajka}

\affiliation{Department of Physics, McGill University, 3600 rue University,
Montreal, Quebec , Canada H3A 2T8}
\affiliation{Institute of Physics, Jan Kochanowski University, Swietokrzyska 15 street, 
25-406 Kielce, Poland}

\author{Sangyong Jeon }

\affiliation{Department of Physics, McGill University, 3600 rue University,
Montreal, Quebec H3A 2T8, Canada}

\date{January 25, 2017}

\begin{abstract}

In this paper we provide a quantum field theoretical study on the shear
and bulk relaxation times. First, we find Kubo formulas for
the shear and the bulk relaxation times, respectively. They are found by examining
response functions of the stress-energy tensor. We use general properties of correlation
functions and the gravitational Ward identity to parametrize analytical
structures of the Green functions describing both sound and diffusion
mode. We find that the hydrodynamic limits of the real parts of the respective
energy-momentum tensor correlation functions provide us with the method of
computing both the shear and bulk viscosity relaxation times. Next, we calculate the
shear viscosity relaxation time using the diagrammatic approach in
the Keldysh basis for the massless $\lambda\phi^4$ theory. We derive a
respective integral equation which enables us to compute $\eta\tau_\pi$
and then we extract the shear relaxation time. 
The relaxation time is shown to be inversely related to the thermal width
as it should be.

\end{abstract}

\pacs{52.27.Ny, 11.30.Pb, 03.70.+k}

\maketitle

\section{Introduction}

Relativistic viscous hydrodynamics seems to be perfectly suited to
investigate and understand collective phenomena characteristic of
strongly interacting matter produced in heavy-ion collisions at the
Relativistic Heavy Ion Collider (RHIC) and the Large Hadron Collider (LHC),
see Refs.~\cite{Gale:2013da,Heinz:2013th} and references therein.
Transport coefficients are inherent ingredients of the hydrodynamic
description. They control the dynamics of a fluid and to find any of the coefficients we
need to employ a microscopic theory.
Due to the multiple-scale nature of the problem, determination of the
full set of the transport coefficients is not a trivial task. 
Since the shear and the bulk
viscosity have already been examined in many papers,
our particular interest is in the kinetic coefficients of the second-order hydrodynamics,
namely the shear and the bulk relaxation times. 
These relaxation times
fix the characteristic time scales
at which the dissipative currents relax to their first-order solutions.
In this study, we use field theoretical approach to investigate the relaxation times.

To understand the microscopic dynamics of plasma constituents and
determine the values of any transport coefficient one needs to use either
kinetic theory or quantum field theory in the weakly coupled limit. Nevertheless, the
quark-gluon plasma, studied experimentally at RHIC and LHC, is
believed to achieve this limit only at sufficiently high temperatures. It was just the
investigation of the shear and bulk viscosities that established
numerous methods of evaluation of transport coefficients in general.
Within the imaginary-time formalism of the
scalar field theory, shear and bulk viscosities were calculated
\cite{Jeon:1992kk,Jeon:1994if} which showed
how to properly
handle the contributing diagrams to obtain the leading order
contribution. In short, in any diagrammatic approach,
the ladder diagrams have to be resummed
to get the leading order transport coefficients
due to the pinching pole effect arising when
small frequency and long-wavelength limits of the correlation function are
taken.

With this knowledge
transport coefficients of QED \cite{Gagnon:2006hi,Gagnon:2007qt} and of
QCD at leading log order \cite{ValleBasagoiti:2002ir} were obtained. The
kinetic theory approach has appeared, in turn, to be very effective, as in the case of
scalar field theory \cite{Jeon:1995zm}, and also successful in determining
transport coefficients of QCD medium in the leading order
\cite{Arnold:2000dr,Arnold:2003zc}. The latter study shows, in
particular, the effectiveness of the kinetic theory to find leading order
results and, at the same time, 
some difficulties to go beyond it \cite{Arnold:2002zm}.

To date, many approaches to the relaxation times have been developed
within kinetic theory. In \cite{York:2008rr} it was shown that the ratio of
the shear viscosity over its relaxation time is proportional to the
enthalpy density and the proportionality factor is slightly different in QCD
than in the $\phi^4$ theory. A similar relation for the ratio was established
within the so-called 14-moment approximation to the Boltzmann equation. And
generally the method of moments, first proposed by Grad and then
further developed by Israel and Stewart, has been examined comprehensively in
\cite{Denicol:2010xn,Denicol:2012es,Denicol:2012cn,Denicol:2014vaa}.
The moment approaches show, in particular, that in order to compute the viscosity
coefficients it is necessary to invert the collision operator and to
determine the relaxation times one has to find the eigenvalues and
eigenvectors of the operator. 

There exist also field-theoretical
studies on the shear relaxation time \cite{Denicol:2011fa,Denicol:2011ef}, where
a general form of a retarded Green function is considered. This work
actually shows the microscopic origin of the shear relaxation time, which is
found to be inversely proportional to the imaginary part of the pole of the
particle propagator.
The Kubo formula engaging shear relaxation time was found for conformal
systems \cite{Baier:2007ix,Moore:2010bu} via the response of a system
to small and smooth perturbations of a background metric. 
The projection operator method was used in \cite{Koide:2009sy} to obtain the shear
relaxation time.

As for the bulk relaxation time,
the projection operator method was 
used in \cite{Huang:2010sa} to obtain it.
Kubo formula for the product of the bulk viscosity and
the bulk relaxation time, $\zeta\tau_\Pi$, can be 
also deduced from response functions studied in Ref.~\cite{Hong:2010at}
using a mixture of the effective kinetic theory and metric perturbations.

Our goal here is to figure out, through a standard formulation of
quantum field theory, how the shear and the bulk relaxation times are related
to microscopic quantities. In order to obtain any transport coefficient
on this ground, one needs to know the corresponding Kubo-type relation.
In this study, by studying both
the sound and the shear modes we are able to find a set of new Kubo-type relations. 
In particular, we find the Kubo formula which relates 
the product $\zeta\tau_\Pi$ to the second derivative
of the real part of the pressure-pressure response function
with respect to the frequency. It is worth emphasizing
that the obtained Kubo relation provides us with an explicit
prescription on how to examine the bulk relaxation time from the quantum 
field theory perspective. All formulas studied here were 
obtained by making use of hydrodynamic limits, general properties of a
response function and the Ward identity. 
The method we employ was introduced in
\cite{Jeon:2015dfa} and here we provide its extension. In the
low-frequency and long-wavelength limits, each of the correlation functions are
related to some set of kinetic coefficients. 

In this paper we focus only on the leading order of the response function, which
is sufficient to study shear effects and, in particular, find the shear
relaxation time. In order to analyze bulk effects one needs to go
to the next-to-leading order calculation, which will be the purpose of
a future work.

To find the value of the shear relaxation time we use diagrammatic
methods of the closed-time path (Keldysh-Schwinger) formalism. As advocated
in \cite{Wang:2002nba}, the $(r,a)$ or Keldysh basis serves very
convenient framework for such considerations, mostly due to the vanishing $aa$
propagator component. We work
with the massless $\lambda\phi^4$ theory in the weakly coupled limit.
We start with the one loop case and then perform the resummation over ladder 
diagrams, which contribute at the same order.
The one-loop approximation allows us to determine what
is the typical scale at which $\eta\tau_\pi$
appears. This is found to be of the order $1/\Gamma_p^2$, where
$\Gamma_p$ is the thermal width and it is directly
related to the mean free path of the system constituents. With
the knowledge on the one-loop result for $\eta$
we are able to extract $\tau_\pi$, which scales as $1/\Gamma_p$. 
The summation of ladder diagrams leads us to
manipulation on the four-point Green functions which couple to each other 
through the Bethe-Salpeter equation. In the $(r,a)$
basis, however, only the $G_{aarr}$ component
matters and consequently the Bethe-Salpeter equation decouples.
To compute the shear viscosity the integral equation for $\text{Im} G_{aarr}$
needs to be solved. We show that to compute $\eta\tau_\pi$ one needs to
solve the integral equation for $\partial_\omega \text{Re} G_{aarr}$, which 
requires us to introduce a new type of the effective vertex. Both $\eta$ 
and $\eta\tau_\pi$ are evaluated numerically to extract the shear 
relaxation time.

The remaining part of the paper is organized as follows. In
Sec.~\ref{sec-hydro-mod} we provide a brief introduction to viscous
hydrodynamics, mainly to write the dispersion relations of the two
hydrodynamic modes arising from the momentum and energy dissipation.
We assume that there are no other currents coupled to the energy-momentum
tensor. In Sec.~\ref{sec-corr-fun} general properties and constraints of the
response functions are discussed. Section~\ref{sec-Kubo-form} presents
the way on how to parametrize the response functions to the longitudinal and
transverse hydrodynamic fluctuations so that to reproduce the
corresponding dispersion relations. Subsequently, we find Kubo-type relations. In
Sec.~\ref{sec-diagram} the expression which allows us to extract the 
relaxation time for shear viscosity is derived
within diagrammatic methods in the real-time formalism. We perform
full leading order analysis providing summation over multi-loop diagrams. 
We also introduce a new effective vertex, which enters the formula for 
evaluation of $\eta\tau_\pi$. In Sec.~\ref{sec-discussion}
we evaluate $\tau_\pi$ and $\langle\epsilon+P\rangle\tau_\pi/\eta$ numerically as 
a function of the constant coupling. We also
discuss our results in the context of kinetic theory findings. 
We conclude in Sec.~\ref{sec-conclusions}.

\section{Hydrodynamic modes}
\label{sec-hydro-mod}

Here, we very briefly introduce basic equations of hydrodynamics,
mostly to fix a starting point for the further discussions. For more comprehensive
analysis we refer the reader to, for example,
\cite{Denicol:2012es,Jeon:2015dfa,Betz:2009zz,Kovtun:2012rj}. The
discussions in this section and Secs.~\ref{sec-corr-fun}
and~\ref{sec-Kubo-form} closely follow \cite{Jeon:2015dfa}.

Hydrodynamics as a long-wavelength and low-frequency many-body effective
theory provides a macroscopic description of a system which is close to
thermal equilibrium. It
governs the evolution of any fluid in terms of
flows of its conserved quantities, such as the energy, momentum or baryon
current. Here, we study phenomena associated only with the
energy-momentum conservation. The corresponding conservation law is the continuity
equation of the energy-momentum tensor $T^{\mu\nu}$
\ba
\label{en-mom-conservation}
\partial_\mu T^{\mu\nu}&=&0.
\ea
When the system is approaching local thermal equilibrium, its
relevant behavior is fairly well described by the viscous
hydrodynamics, for which the energy-momentum tensor takes the form
\ba
\label{vis-T}
T^{\mu\nu} = \epsilon u^\mu u^\nu - \Delta^{\mu\nu}(P+\Pi) +
\pi^{\mu\nu},
\ea
where $\epsilon$ is the energy density, $P$ is the thermodynamic pressure, $u^\mu$ are
the components of the flow velocity with the normalization condition
$u^\mu u_\mu=1$, $\Delta^{\mu\nu} = g^{\mu\nu} - u^\mu u^\nu$ is the
projection operator with $u_\mu \Delta^{\mu\nu} = 0$,
and the Minkowski metric is 
$g^{\mu\nu}=(1,-1,-1,-1)$. The terms
$\Pi$ and $\pi^{\mu\nu}$ are the bulk viscous pressure and the shear
stress tensor, respectively. They are viscous corrections which contain
the dynamics of the dissipative medium approaching the
equilibrium state. The shear tensor is symmetric, traceless, $\pi^\mu_\mu=0$, and
transverse, $u_\mu\pi^{\mu\nu}=0$, and the bulk pressure $\Pi$ is a
correction to the thermodynamic pressure. These corrections are
assumed to be small for the nearly equilibrium state of a system. If the static
equilibrium limit is achieved, the dissipative corrections in Eq.~(\ref{vis-T})
vanish and one reproduces the ideal hydrodynamic stress-energy tensor. Within the
Navier-Stokes approach, the viscous corrections are
obtained from the gradient expansion of the energy density and flow velocity.
Then, up to the linear terms, only the corrections proportional to
$\partial^\mu u^\nu$ matter. The dissipative currents $\Pi$ and $\pi^{\mu\nu}$ take
the following forms
\ba
\label{bulk-viscosity-NS}
\Pi_{\textrm{NS}} &=& \zeta \Delta_{\mu\nu} \partial^\mu u^\nu,\\
\label{shear-viscosity-NS}
\pi^{\mu\nu}_{\textrm{NS}}&=& 2\eta \Delta^{\mu\nu}_{\alpha\beta}
\partial^\alpha u^\beta,
\ea
where $\zeta$ and $\eta$ are the bulk and shear viscosities which
determine transport phenomena of the energy and momentum, and
$\Delta^{\mu\nu}_{\alpha\beta} \equiv (\Delta^\mu_\alpha
\Delta^\nu_\beta +
\Delta^\mu_\beta \Delta^\nu_\alpha
-2/3\Delta^{\mu\nu}\Delta_{\alpha\beta})/2$ is the traceless and
transverse projection operator. In the fluid cell rest frame the dissipative
currents may be written as
\ba
\label{bulk-viscosity-NS-ij}
\Pi_{\textrm{NS}} &=& - \gamma \partial_l T^{l0},\\
\label{shear-viscosity-NS-ij}
\pi^{ij}_{\textrm{NS}}&=& D_T \bigg( \partial^i T^{j0} + \partial^j
T^{i0}
-\frac{2}{3} g^{ij}\partial_l T^{l0} \bigg), 
\ea
where $\gamma=\zeta/(\epsilon+P)$ and $D_T=\eta/(\epsilon+P)$.
Accordingly, the full spatial viscous
correction to the energy-momentum tensor of a viscous fluid is
\ba
\label{full-corr}
\delta T^{ij} = \pi^{ij}_{\textrm{NS}}+\Pi_{\textrm{NS}}= D_T \bigg(
\partial^i T^{j0} + \partial^j T^{i0} -\frac{2}{3} g^{ij}\partial_l T^{l0} \bigg) 
+g^{ij} \gamma \partial_l T^{l0}.
\ea
in the rest frame of the fluid cell at the position $x$.

In fact, the Navier-Stokes theory of a relativistic fluid is
acausal and unstable. Generally speaking, the viscous currents must
be allowed to take some time when responding to changes in the thermodynamic forces. 
This requires taking the next terms in gradient expansion, 
and the corresponding hydrodynamics is the second-order Israel-Stewart theory.
The bulk viscous pressure and the shear stress tensor
are then subject to the relaxation equations. 
In a general frame, these are given by
\ba
\label{rel-time-shear}
\pi^{\mu\nu}&=&\pi^{\mu\nu}_{NS} - \tau_\pi \dot \pi^{\langle \mu\nu
\rangle},\\
\label{rel-time-bulk} 
\Pi&=&\Pi_{NS} - \tau_\Pi \dot \Pi,
\ea
where we ignored the non-linear terms as they are not relevant for our study.
We have also used the notation $A^{\langle \mu\nu \rangle} \equiv
\Delta^{\mu\nu}_{\alpha\beta}A^{\alpha\beta}$ for the spin-2 component of a 
rank-2 tensor.
The new transport
coefficients, $\tau_\Pi$ and $\tau_\pi$, are relaxation times for the
bulk and shear viscosities, respectively. They determine how fast the bulk
pressure and the shear tensor relax to the respective Navier-Stokes
forms given by Eqs.~(\ref{bulk-viscosity-NS}) and (\ref{shear-viscosity-NS}),
respectively. As a result, causality of the theory is maintained if
the relaxation times satisfy certain restrictions \cite{Pu:2009fj}. The
relaxation equations take the following forms in the local rest frame
\ba
\label{relax-eq-11}
\partial_t \Pi &=& -\frac{\Pi -\Pi_{\textrm{NS}}}{\tau_\Pi}, \\
\label{relax-eq-22}
\partial_t \pi^{ij} &=&
-\frac{\pi^{ij}-\pi^{ij}_{\textrm{NS}}}{\tau_\pi}.
\ea

If there are no other currents coupled to the energy-momentum tensor,
there are two hydrodynamic modes that determine the behavior of the system.
These are the diffusion and sound modes. The diffusion mode describes fluid
flow in the direction transverse to the flow velocity. It appears as a
consequence of the momentum conservation, which is
\ba
\label{mom-cons}
\partial_t T^{0k}=-\partial_l T^{lk}.
\ea
When the relaxation equation (\ref{relax-eq-22}) and the Navier-Stokes form
of the shear tensor (\ref{shear-viscosity-NS-ij}) are implemented to
the momentum conservation law (\ref{mom-cons}), we get the corresponding
hydrodynamic equation, which is the equation of motion of the
transverse part of stress tensor
\ba
\label{eom-mom}
0 &=& (\tau_\pi \partial_t^2+\partial_t-D_T \nabla^2)\pi_T^i,
\ea
where $\pi^i_T=\epsilon_{ijk}\partial_j T^{k0}$. The corresponding
dispersion relation is then found to be
\ba
\label{disp-rel-mom}
0 &=& -\omega^2 \tau_\pi -i \omega +D_T {\bf k}^2
\ea
with $\omega$ and ${\bf k}$ being the frequency and wavevector of the
momentum diffusion excitation. 

The other mode is associated with
small disturbances in dynamic variables propagating longitudinally in
the medium. The conservation law in the local rest frame then is
\ba
\label{en-con}
\partial_t^2 \epsilon &=& \nabla^2 P - \partial_l\partial_m \pi^{lm}
+\nabla^2 \Pi.
\ea
By multiplying Eq.~(\ref{en-con}) by $(\tau_\pi \partial_t+1)(\tau_\Pi\partial_t+1)$, 
making use of the relaxation equations (\ref{relax-eq-11}) and (\ref{relax-eq-22}) 
and using the Navier-Stokes forms of the stress
tensor (\ref{shear-viscosity-NS-ij}) and bulk pressure
(\ref{bulk-viscosity-NS-ij}), we get the equation of motion for the energy density
deviation $\delta \epsilon$,
\ba
\label{eom-en}
0 &=& \Big[\partial_t^2 - v_s^2 \nabla^2 + (\tau_\pi + \tau_\Pi)
\partial_t^3
- (\tau_\pi + \tau_\Pi)v_s^2 \nabla^2 \partial_t  
- \frac{4D_T}{3} \nabla^2 \partial_t  
-\gamma \nabla^2 \partial_t
\nn \\ 
&&
+
\tau_\pi \tau_\Pi \partial_t^4 -\tau_\pi \tau_\Pi v_s^2 \nabla^2
\partial_t^2  
- \frac{4D_T}{3} \tau_\Pi \nabla^2 \partial_t^2 
- \gamma \tau_\pi \nabla^2 \partial_t^2  \Big] \delta \epsilon,
\ea
where $v_s^2=\partial P/\partial \epsilon$ is the speed of sound. The
solution to Eq.~(\ref{eom-en}) is provided by the following
dispersion relation:
\ba
\label{disp-rel-en}
0&=& - \omega^2 + v_s^2 {\bf k}^2
+ i\omega^3 (\tau_\pi + \tau_\Pi)
- i\bigg(\frac{4D_T}{3}+\gamma+v_s^2(\tau_\pi + \tau_\Pi) \bigg)\omega{\bf
k}^2 
\\ \nn
&&
+\tau_\pi \tau_\Pi \omega^4 - \tau_\pi \tau_\Pi v_s^2 \omega^2 {\bf
k}^2
- \tau_\Pi \frac{4D_T}{3} \omega^2 {\bf k}^2 
- \tau_\pi \gamma \omega^2 {\bf k}^2 .
\ea
The dispersion relations (\ref{disp-rel-mom}) and (\ref{disp-rel-en})
play an essential role in further analysis as they encode full information on
the relaxation times for viscosities. This information should be also
contained in the pole structure of the respective retarded Green function and
this is the subject of the next sections. 

\section{Response functions}
\label{sec-corr-fun}

Linear response theory is a natural quantum-mechanical framework to
examine systems exhibiting small deviations from equilibrium. Within the
linear response theory one is able to express quantities
characteristic of the nonequilibrium state of a fluid in terms of time dependent
correlation functions of the equilibrium state. The linear response theory is explained in many
textbooks, see for example \cite{Kapusta:2006pm}, and here we
restrict ourselves to discuss only those properties of response functions relevant
to our study of transport coefficients.

When a system undergoes small perturbations, the deviation of an
observable $A$ from equilibrium is encoded in equilibrium response function as
\ba
\label{A-obs}
\delta \langle \hat A(t, {\bf x}) \rangle = \int d^4 x' G_R (t-t',
{\bf x}
- {\bf x'}) \theta(-t') e^{\varepsilon t'} f({\bf x}),
\ea
where $t>0$, $\theta(t)$ is the standard step function, $f({\bf x})$ is
an
external perturbing force coupled to $\langle \hat A \rangle$ and acts on the system with
infinitesimally slow rate $\varepsilon$, and $\langle \cdots \rangle$ means the
thermal expectation value. $G_R$ is the retarded response function
corresponding to the Hermitian operator $\hat A$,
\ba
\label{ret-corr-fun}
G_R (t-t', {\bf x} - {\bf x'}) =- i \theta(t-t') \langle [\hat
A_H(t,{\bf
x}), \hat A_H(t',{\bf x}')] \rangle,
\ea
where $ \hat A_H$ stands for the operator in the Heisenberg picture.
For further analysis it is also convenient to introduce the advanced
correlation function, which is
\ba
\label{adv-corr-fun}
G_A (t-t', {\bf x} - {\bf x'}) = i \theta(t'-t) \langle [\hat
A_H(t,{\bf
x}), \hat A_H(t',{\bf x}')] \rangle.
\ea

Suppose the retarded Green function $G_R$ satisfies the following equation of motion:
\ba
\label{op-D}
D_A G_R (t-t', {\bf x} - {\bf x'}) = 
d_A \delta(t-t') \delta({\bf x} -{\bf x}'),
\ea
where $D_A$ is some operator such that $G_R$ is its generalized 
Green function and $d_A$ may contain
a finite number of derivatives. For positive values of $t$, $t\ne t'$ since $t'$ is
restricted by $\theta(-t')$.
Therefore, for positive $t$, 
$\delta \langle \hat A \rangle$ satisfies the following evolution equation:
\ba
\label{ev-eq}
D_A \delta \langle \hat A(t,{\bf x}) \rangle=0.
\ea
Accordingly, the evolution equation is known whenever one finds the pole structure of
the response function \cite{Kadanoff}. 

The formula (\ref{A-obs}) shows explicitly that the linear response of the
system is expressed in terms of a retarded Green function of
Heisenberg operators. To study the retarded Green functions, it is convenient 
to introduce the spectral density defined by the thermal expectation value
of the commutator
\ba
\label{spectral-density}
\rho^{AA}(k) = \int d^4 x e^{ikx} \langle [\hat A_H(x), \hat A_H(0)]
\rangle,
\ea
where $k=(\omega, {\bf k})$, which may be expressed as
\ba
\label{spectral-density-2}
\rho^{AA}(k) = \frac{1}{Z_0} \sum_{m,n} \big( e^{-\beta E_n} -
e^{-\beta
E_m} \big) 
(2\pi)^4 \delta(k-p_m+p_n) \big| \langle p_n |\hat A | p_m \rangle
\big|^2,
\ea
when $\hat A$ is Hermitian. Here $|m \rangle$ is the simultaneous eigenstate of the system's
total Hamiltonian $\hat H$ and the total momentum $\hat{\bf P}$
with the eigenvalue $p_m=(E_m,{\bf p}_m)$. Relying on the fact that
for any observable the corresponding operator must be Hermitian, 
$\hat A^\dagger =\hat A$, one can derive
\ba
\rho^{AA}(-\omega,-{\bf k})=-\rho^{AA}(\omega, {\bf k}). 
\ea
For an equilibrium system, which is isotropic, $\rho^{AA}(\omega,{\bf k})$
must preserve a rotational invariance so that it depends on momentum
only through its absolute value $|{\bf k}|$. Therefore, the spectral
density is an odd function of $\omega$, that is, $\rho^{AA}(-\omega, {\bf
k})=-\rho^{AA}(\omega,{\bf k})$.

In the spectral representation the retarded and advanced Green
functions are given by
\ba
\label{spectral-GF}
G_{R/A}(\omega,{\bf k})= \int \frac{d\omega'}{2\pi} 
\frac{\rho^{AA}(\omega',{\bf k})}{\omega'-\omega \mp i\epsilon},
\ea
where the upper sign $(-)$ corresponds to the retarded function and the
lower sign $(+)$ to the advanced one. By extracting the principal value of
the integral in Eq. (\ref{spectral-GF}) from the imaginary part one obtains
the following relations:
\ba
\label{real-part-spectral}
\textrm{Re}\;G_R(\omega, {\bf k}) &=& \textrm{Re}\;G_A(\omega, {\bf
k})=
\mathcal{P} \int \frac{d\omega'}{2\pi} \frac{\rho^{AA}(\omega',{\bf
k})}{\omega'-\omega}, \\
\label{imaginary-part-spectral}
\textrm{Im}\;G_R(\omega, {\bf k}) &=& -\textrm{Im}\;G_A(\omega, {\bf
k}) =
\frac{1}{2}\rho^{AA}(\omega,{\bf k}),
\ea
where $\mathcal{P}$ stands for the principal value. By changing the
sign of $\omega'$ in the formula (\ref{real-part-spectral}) and using the fact that the
spectral function is an odd function of the frequency one observes
that $\textrm{Re}\,G_R(\omega, {\bf k})=\textrm{Re}\,G_R(-\omega, {\bf
k})$, that is, the real part of the retarded and the advanced Green
function is an even function of frequency. Moreover, the imaginary
part of the retarded response function, since related directly to the
spectral function, is an odd function of $\omega$. These facts will be
frequently used in the next parts of this paper.

Due to the fact that the stress-energy tensor represents both the conserved current as well
as the generators of the space time evolution, the correlation functions of $T^{\mu\nu}$
are not so simple. To determine them correctly, one must first start with the following 
gravitational Ward identity \cite{Deser:1967zzf}:
\ba
\label{Ward-id-coord}
\partial_\alpha \big[ \bar G^{\alpha\beta,\mu\nu}(x,x') 
- \delta^{(4)}(x-x') \big(g^{\beta\mu} \langle \hat T^{\alpha\nu}(x')
\rangle 
+ g^{\beta\nu} \langle \hat T^{\alpha\mu}(x') \rangle -
g^{\alpha\beta}
\langle \hat T^{\mu\nu}(x') \rangle \big) \big]=0,
\ea
which becomes in the momentum space
\ba
\label{Ward-id-real}
k_\alpha \big( \bar G^{\alpha\beta,\mu\nu}(k) 
- g^{\beta\mu} \langle \hat T^{\alpha\nu} \rangle 
- g^{\beta\nu} \langle \hat T^{\alpha\mu} \rangle 
+ g^{\alpha\beta} \langle \hat T^{\mu\nu} \rangle \big)=0,
\ea
where $k=(\omega, {\bf k})$. The identity (\ref{Ward-id-coord}) is most conveniently 
derived in the imaginary-time metric. 
The two-point functions of $T^{\mu\nu}$ is then obtained by taking the second
 functional derivative of the partition
function with respect to the imaginary-time metric.
Going to the flat space and then analytic continuing to the real space give
(\ref{Ward-id-coord}). Let us stress that $\bar
G_R^{\mu\nu,\alpha\beta}(x,x')$ is the response function, which is
{\it not}, in general, the same as
\ba
G_R^{\mu\nu,\alpha\beta}(x,x')=
-i\theta(x_0-x_0') \langle [\hat T^{\mu\nu}(x), \hat T^{\alpha\beta}(x')] \rangle
\ea
due to the presence of the single stress-energy tensor average terms
in Eq.~(\ref{Ward-id-real}). $\bar G_R$ differs  from $G_R$ by terms
containing $\delta(x-x')$. Let us add that the formula (\ref{Ward-id-real}),
when combined with the continuity equations, fixes actually a set of
constraints that the response functions, corresponding to different
components of the energy-momentum tensor, must maintain.

\section{Analytic structure of response functions and Kubo formulas}
\label{sec-Kubo-form}

The general properties of a correlation function supported by the
Ward identity constrain its analytical form 
enough so that one is able to parametrize it 
for both the propagating and diffusive mode. Here, we
parametrize the correlation functions in terms of the linear response
method.

\subsection{Response function to transverse fluctuations}
\label{subsec-transverse}

The perturbing Hamiltonian for the shear flow is
\ba
\delta \hat H(t) = -\int d^3 x \theta(-t) e^{\varepsilon t} \hat T^{x0}(t,{\bf x}) \beta_x(y)
\ea
Note that the external force $\beta_x(y)$ is related to the flow velocity component
in the $x$-direction which only varies in the perpendicular $y$-direction. Hence,
at $t = 0$, this sets up a system with a non-zero shear flow.
The corresponding linear response is
\ba
\label{pert-ham-shear}
\delta \langle \hat T^{x0}(t,k_y) \rangle = 
\beta_x(k_y) \int_{-\infty}^\infty dt' \theta(-t') e^{\varepsilon t'} 
\bar G_R^{x0,x0}(t-t',k_y)
\ea
for $t>0$. From the Ward identity (\ref{Ward-id-real}), one finds
\ba
\label{from-ward}
\omega \big( \bar G_R^{x0,x0}(\omega, k_y) + \epsilon \big) &=& 
k_y \bar G_R^{x0,xy}(\omega,k_y), \\
\omega \bar G_R^{x0,xy}&=&k_y \big( \bar G_R^{xy,xy}(\omega,k_y) + P \big).
\ea
When these two equations are combined,
one gets
\ba
\label{two-eq}
\bar G_R^{xy,xy}(\omega,k_y) + P = 
\frac{\omega^2}{k_y^2} 
\big( \bar G_R^{x0,x0}(\omega,k_y) + \epsilon \big).
\ea
In the $\omega \rightarrow 0$ limit, $\bar G_R^{x0,x0}(\omega,k_y)$
must have a well defined limit since it is a thermodynamic quantity.
Moreover, both correlation functions must be well behaved in the
$k_y \rightarrow 0$ limit. Using these arguments 
and the fact that the imaginary part of the retarded Green function must be
an odd function of $\omega$,
one can parametrize $\bar G_R^{xy,xy}(\omega,k_y)$ as
\ba
\label{G-xyxy}
\bar G_R^{xy,xy}(\omega,k_y) = 
\frac{\omega^2[\epsilon + g_T(k_y) + i\omega A(\omega,k_y)]}
{k_y^2 - \frac{i\omega}{D(\omega,k_y)}-\omega^2 B(\omega,k_y)} - P,
\ea
and
\ba
\label{G-x0x0}
\bar G_R^{x0,x0}(\omega,k_y) = 
\frac{k_y^2[\epsilon + g_T(k_y) +i\omega A(\omega,k_y)]}{k_y^2 - \frac{i\omega}
{D(\omega,k_y)}-\omega^2 B(\omega,k_y)} - \epsilon,
\ea
where $g_T(k_y)=\bar G^{x0,x0}(0,k_y) = P + g_{\pi\pi}(k_y)$ comes
from Eq.~(\ref{eq:app_pi_pi}) in Appendix \ref{app:ward_id}.
The functions $A$, $B$, and $D$ have the form
\ba
\label{functions}
D(\omega,k_y)=D_R(\omega,k_y)-i\omega D_I(\omega,k_y),
\ea
where $D_R(\omega,k_y)$ and $D_I(\omega,k_y)$ are real-valued even
functions of $\omega$ and $k_y$. The real parts $D_R$ and $B_R$ must
have a non-zero limit as $\omega \rightarrow 0$ and $k_y \rightarrow 0$. All
other parts of $A$, $B$, and $D$ must have finite limits as $\omega
\rightarrow 0$ and $k_y \rightarrow 0$.
Dynamical information in the hydrodynamic limit is contained in the
constants $D_R(0,0)$, $D_I(0,0)$, $B_R(0,0)$, etc.

The pole structure of a Green function determines the corresponding
dispersion relation of a given hydrodynamic mode. Therefore, the dispersion
relation of the diffusive excitation is dictated by the form of the denominator of
the function (\ref{G-xyxy}), that is
\ba
\label{diff-mode}
k_y^2 D_R - i\omega k_y^2 D_I - i\omega -\omega^2B_R D_R 
+i\omega^3 B_I D_R + i\omega^3 B_R D_I +\omega^4 B_I D_I=0.
\ea

By comparing the pole structure (\ref{diff-mode}) to the dispersion
relation obtained from the conservation law (\ref{disp-rel-mom}), we
find the following relations
\ba
D_R(0,0)&=&D_T, \\ 
B_R(0,0)&=&\frac{\tau_\pi}{D_T}.
\ea
From this, one finds
\begin{equation}
\bar{G}_R^{xy,xy}(\omega,0)
= i\omega \eta 
-\eta\tau_\pi\omega^2 + 
(\epsilon + P)
\left( D_I(0,0) +
 A_R(0,0)\eta \right)\omega^2 + \mathcal{O}(\omega^3)
\end{equation}
Hence,  when the small $\omega$ and $k_y$ limits of the function
(\ref{G-xyxy}) are taken, we find
\ba
\label{KF-1r}
\eta = \lim_{\omega \rightarrow 0} \lim_{k_y \rightarrow 0}
\frac{1}{\omega} \text{Im}\bar
G_R^{xy,xy}(\omega, k_y),
\ea
which is the Kubo relation for the shear viscosity. The
hydrodynamic limits also enable us to find
\ba
\label{KF-2r}
\eta \tau_\pi - (\epsilon+P)\left( D_I(0,0) +
A_R(0,0)\eta \right)
= -\frac{1}{2}\lim_{\omega \rightarrow 0} \lim_{k_y \rightarrow
0}\partial^2_\omega \;
\text{Re} \bar G^{xy,xy}_R(\omega, k_y),
\ea
where we have used $P=g_T(0,0)$. 

In the relation (\ref{KF-2r}), there
appear the constants $A_R(0,0)$ and $D_I(0,0)$, which we are not able
to identify within this approach. However, other studies involving a slightly different perturbing Hamiltonian
were able to identify the second term in (\ref{KF-2r}) as a thermodynamic quantity.
The Kubo relations for the second-order hydrodynamics coefficients
were examined in \cite{Baier:2007ix,Moore:2010bu,Moore:2012tc}, where they
are provided by studying the response of a fluid to small and smooth metric
perturbations. If one takes into account only linearized equations, the following
relations are found:\footnote{In the formulas (\ref{eta-GM})--(\ref{corr-fun-GM}) we applied the 
same sign convention as used in the entire paper, that is, with the metric being mostly negative. 
In the original papers \cite{Baier:2007ix,Moore:2010bu,Moore:2012tc} these formulas are given 
with the opposite sign convention since they are studied in the flat space which is convenient
when one examines transport properties of a medium via background geometry perturbations.}
\ba
\label{eta-GM}
\eta &=& i \lim_{\omega \rightarrow 0} 
\lim_{k_z \rightarrow 0} \bar G_R^{xy,xy}(\omega,k_z), \\
\label{kappa_def}
\kappa&=&  
\lim_{k_z \rightarrow 0} \lim_{\omega \rightarrow 0}
\partial_{k_z}^2 \bar G_R^{xy,xy}(\omega,k_z),\\
\label{eta-tau-GM}
\eta \tau_\pi&=& -\frac{1}{2} \lim_{\omega \rightarrow 0} \lim_{k_z
\rightarrow 0}
\partial_\omega^2  \bar G_R^{xy,xy}(\omega,k_z) 
+ \frac{1}{2} \lim_{k_z \rightarrow 0} \lim_{\omega \rightarrow 0}
\partial_{k_z}^2 \bar G_R^{xy,xy}(\omega,k_z) ,
\ea
where $\kappa$ is an additional coefficient. Note that these formulas involve non-zero
$k_z$ while our formulas involve $k_y$. When expanded  
around small $\omega$ and $k_z=0$,
the correlation function $\bar G_R^{xy,xy}(\omega,k_z)$ becomes
\ba
\label{corr-fun-GM}
\bar G_R^{xy,xy}(\omega,0) \approx -P+i\omega \eta -\eta\tau_\pi \omega^2
+\frac{\kappa}{2}\omega^2 .
\ea
Since the two-point functions (\ref{G-xyxy})
and (\ref{corr-fun-GM}) have different momentum arguments,
their analytical structures are slightly different. See Appendix \ref{app:decomp} for details.
As pointed out in \cite{Baier:2007ix}, the $\kappa$ term in the
dissipative part of the stress-momentum tensor is proportional to $u^\mu$ so that
these additional terms in the correlation function do not come from the
contact term in the coordinate space. Nevertheless, providing
$\bar G_R^{xy,xy}(\omega,k_z)$ and $\bar G_R^{xy,xy}(\omega,k_y)$ share the same
diffusion pole structure, their small frequency and {\it vanishing}
momentum limits should be consistent with each other.

In the small $\omega$ and vanishing $k_y$ limits, the function
(\ref{G-xyxy}) is 
\ba
\label{lim-G-xyxy}
\bar G_R^{xy,xy}(\omega,0) &\approx& 
-P + i\omega(\epsilon +P)D_R(0,0) \\ \nn
&&+\omega^2\big[-(\epsilon+P)D_R^2(0,0)B_R(0,0)
+(\epsilon+P)D_I(0,0)+A_R(0,0)D_R(0,0) \big].
\ea
By comparing the function (\ref{lim-G-xyxy}) to (\ref{corr-fun-GM})
 we can obtain the condition on the unknown functions $A_R(0,0)$ and
$D_I(0,0)$ and clarify the relation (\ref{KF-2r}). So we identify
\ba
\eta&=& (\epsilon +P)D_R(0,0), \\ \nn
\eta\tau_\pi&=&\eta D_R(0,0)B_R(0,0).
\ea
Then the condition on the contribution from $D_I(0,0)$ and $A_R(0,0)$
is
\ba
\label{cond-AD}
(\epsilon+P)D_I(0,0) 
+
A_R(0,0) D_R(0,0) = \frac{\kappa}{2}.
\ea
Finally, we can write the Kubo relation for $\eta\tau_\pi$ as 
\ba
\label{Kubo-eta-tau-shear}
\eta \tau_\pi - \frac{\kappa}{2}
&=& 
-\frac{1}{2} \lim_{\omega \rightarrow 0} \lim_{k_y \rightarrow 0}
\partial_\omega^2 \;
\textrm{Re} \bar G_R^{xy,xy}(\omega,k_y).
\ea
It is known \cite{Romatschke:2009ng,Moore:2012tc} that
$\kappa = \mathcal{O}(\lambda^0 T^2)$ in the weak coupling limit.
On the other hand, both $\eta$ and $\tau_\pi$ behave like the mean free path
which
depends inversely on the cross section. Hence,  Eq. (\ref{Kubo-eta-tau-shear}) 
can be still used to study the leading order shear relaxation time.

\subsection{Response function to longitudinal fluctuations}
\label{subsec-longitudinal}

The perturbing Hamiltonian for the bulk flow is
\ba
\label{ham-pert-en}
\delta \hat H(t) = 
- \int d^3x \theta(-t) \hat T^{00}(x) e^{\varepsilon t} \beta_0({\bf x}),
\ea
where $\hat T^{00}$ is the operator of energy density and $\beta_0({\bf x})$ 
is an space-dependent external force, which has driven the system off
equilibrium. The response of the medium then is
\ba
\label{response-en}
\langle T^{00}(t,{\bf k}) \rangle = 
\beta_0({\bf k}) \int_{-\infty}^\infty dt' \theta(-t') e^{\varepsilon t'} 
\bar G_R^{00,00}(t-t',{\bf k}).
\ea
By applying the continuity equation to each index of $\bar{G}_R^{\alpha\beta, \mu\nu}$ 
in the Ward identity $(\ref{Ward-id-real})$, we get
\ba
\label{W-id-en-dist}
\omega^4 \bar{G}_R^{00,00}(\omega, {\bf k}) = 
\omega^4 \epsilon - \omega^2 {\bf k}^2 (\epsilon +P) 
+ {\bf k}^4 \bar{G}_L (\omega, {\bf k}),
\ea
where $\bar{G}_L (\omega, {\bf k})$ is the response function to the
longitudinal fluctuations and, through the Ward identity, it is
related to the spatial stress-stress function $\bar G_R^{ij,mn}(\omega,{\bf k})$
as
\ba
\label{GL}
{\bf k}^4 \bar{G}_L (\omega, {\bf k}) = k_i k_j k_m k_n [\bar
G_R^{ij,mn}(\omega,{\bf k})
+P(\delta^{im}\delta^{jn}+\delta^{in}\delta^{jm}-\delta^{ij}\delta^{mn})].
\ea
What is more, the Ward identity enables one to express a response function
associated with an arbitrary energy-momentum tensor component via the
spatial stress-stress response function $\bar G_R^{ij,mn}(\omega,{\bf k})$.

{}From the response function $\bar{G}_L$ (\ref{W-id-en-dist}), one finds
\ba
\label{GL-small}
\bar{G}_L (\omega, {\bf k}) \approx 
\frac{\omega^2}{{\bf k}^2} (\epsilon +P) + \frac{\omega^4}{{\bf k}^4}
\big( \bar G_R^{00,00}(0,{\bf k}) -\epsilon \big).
\ea
We know that $\bar{G}_R^{00,00}(0,{\bf k}) = Tc_v+ O(k^2)$ where $c_v$ is the
specific heat per unit volume from Eq. (\ref{eq:app_c_v}) in Appendix \ref{app:ward_id}.
It is also related to the speed of sound $Tc_v = (\epsilon + P)/v_s^2$.
We also take into account that the imaginary part of $\bar G_L$ must
be an odd function of $\omega$ and the full function should
behave well in the ${\bf k}\to 0$ limit.
All these arguments allow one to
parametrize the most general form
 of the function $\bar G_L$ as \cite{Jeon:2015dfa}\footnote{
In \cite{Jeon:2015dfa}, the numerator had $i\omega^3 Q(\omega, {\bf k})$
instead of $\omega^2Q(\omega,{\bf k})$.
Equation (\ref{GL-gen-form}) is the correct form for the most general parametrization.
} 
\ba 
\label{GL-gen-form}
\bar G_L(\omega,{\bf k})=
\frac{\omega^2[\epsilon+ P  +\omega^2 Q(\omega, {\bf k}) ]}
{{\bf k}^2 - \frac{\omega^2}{Z(\omega, {\bf k})} 
+ i\omega^3R(\omega, {\bf k})}.
\ea
The functions $Z(\omega,{\bf k})$, $R(\omega,{\bf k})$, and $Q(\omega,{\bf k})$ 
are all of the form
\be
\label{functions-form}
Z(\omega,{\bf k})=Z_R(\omega,{\bf k}) - i \omega Z_I(\omega,{\bf k}),
\ee
where $Z_R(\omega,{\bf k})$ and $Z_I(\omega,{\bf k})$ are real-valued
even functions of $\omega$ and ${\bf k}$. The real parts $Z_R$ and $R_R$
must have a non-zero limit as $\omega \rightarrow 0$ and ${\bf k}
\rightarrow 0$. All other parts of $Z$, $Q$, and $R$ must have finite limits as
$\omega \rightarrow 0$ and ${\bf k} \rightarrow 0$. 

The pole structure of the correlation function (\ref{GL-gen-form})
provides us with the dispersion relation
\ba 
\label{poles-GL}
\omega^2 -{\bf k}^2 Z(\omega,{\bf k}) - i\omega^3R(\omega,{\bf k})Z(\omega,{\bf k})=0.
\ea
Comparing the dispersion relation (\ref{poles-GL}) to (\ref{disp-rel-en})
one observes that in order to reproduce terms $\sim \omega^2 {\bf k}^2$ and
other terms of higher powers, it is enough to expand 
the real part of $Z(\omega,{\bf k})$ up to $\omega^2$ so that
\ba
\label{Z-r}
Z_R(\omega,{\bf k})= Z_{R1}(0,0) - \omega^2 Z_{R2}(0,0) 
+ \mathcal{O}({\bf k}^2)
+ \mathcal{O}(\omega^4).
\ea
Then, the expression~(\ref{poles-GL}) takes the form
\ba 
\label{poles-GL-A-2}
0&=& \omega^4[Z_I(0,0) R_R(0,0) + R_I(0,0)
Z_{R1}(0,0)] 
- \omega^2 {\bf k}^2 Z_{R2}(0,0) \\ \nn
&& - \omega^2 + {\bf k}^2Z_{R1}(0,0) -i\omega {\bf k}^2
Z_I(0,0) 
+ i\omega^3R_R(0,0) Z_{R1}(0,0) \\ \nn
&&
+ \mathcal{O}(\omega^5) + \mathcal{O}({\bf k}^4).
\ea
Comparing this pole structure to the dispersion relation provided by
a purely hydrodynamic framework (\ref{disp-rel-en}) one finds
the following relations
\ba
\label{coef}
Z_{R1}(0,0) &=& v_s^2,\\
Z_{R2}(0,0) &=& \tau_\pi \tau_\Pi v_s^2 + \tau_\Pi \frac{4D_T}{3} +
\tau_\pi \gamma,\\
Z_I(0,0) &=& \frac{4D_T}{3}+\gamma+v_s^2(\tau_\pi + \tau_\Pi),\\
R_R(0,0) &=& \frac{\tau_\pi + \tau_\Pi}{v_s^2},\\
\label{coef-2}
R_I(0,0) &=& \frac{v_s^2\tau_\pi \tau_\Pi - v_s^2(\tau_\pi +
\tau_\Pi)^2 -
(4D_T/3 + \gamma)(\tau_\pi + \tau_\Pi)}{v_s^4}.
\ea

The imaginary and real parts of the Green function
$\bar{G}_L(\omega,{\bf k})$ in the thermodynamic limit
becomes
\ba
\label{KF-1}
\lim_{\omega \rightarrow 0} \lim_{{\bf k} \rightarrow 0}
\frac{1}{\omega}\text{Im }\bar{G}_L(\omega, {\bf k})
&=&(\epsilon+P)(Z_I-Z_{R1}^2 R_R), \\
\label{KF-2}
-\frac{1}{2} \lim_{\omega \rightarrow 0} \lim_{{\bf k} \rightarrow 0}
\partial^2_\omega \text{Re } \bar{G}_L(\omega, {\bf k})
&=& (\epsilon+P) \Big(2 R_R Z_I Z_{R1} +R_I Z_{R1}^2 - R_R^2 Z_{R1}^3
- Z_{R2} \Big) + Q_R Z_{R1},
\nonumber\\
\ea
where all the constants $Z_I$, $Z_{R1}$, $Z_{R2}$, $R_I$, $R_R$,
and $Q_R$ should be understood as $Z_I \equiv Z_I(0,0)$, etc.
Using the relations (\ref{coef})--(\ref{coef-2}), we find the following Kubo
formulas
\ba
\label{Kubo-GR}
\frac{4 \eta}{3} + \zeta 
&=& 
\lim_{\omega \rightarrow 0} \lim_{{\bf k}\rightarrow 0}
\frac{1}{\omega}\text{Im }\bar{G}_L(\omega, {\bf k}), \\
\label{Kubo-rel-2}
\frac{4}{3} \eta \tau_\pi + \zeta \tau_\Pi  + Q_R v_s^2
&=& 
-\frac{1}{2} \lim_{\omega \rightarrow 0} \lim_{{\bf k} \rightarrow 0} 
\partial^2_\omega \text{Re} \bar{G}_L(\omega, {\bf k}), \\
\label{Kubo-rel-vs}
v_s^2 (\epsilon + P)
&=& 
- \lim_{\omega \rightarrow 0} \lim_{{\bf k} \rightarrow 0} \bar{G}_L(\omega, {\bf k}).
\ea
Here as in the shear case, one constant, $Q_R$, is left undetermined
through this analysis. In \cite{Hong:2010at}, it was shown through the curved metric
analysis that
\be
\frac{4}{3} \eta \tau_\pi + \zeta \tau_\Pi -{2\kappa\over 3}
=
-\frac{1}{2} \lim_{\omega \rightarrow 0} \lim_{{\bf k} \rightarrow 0}
\partial^2_{\omega} \textrm{Re} \bar G_L (\omega, {\bf k}).
\ee
Hence, we may identify 
\be
Q_R = -{2\kappa\over 3v_s^2}.
\ee

More convenient Kubo formulas can be obtained if one combines 
the Kubo formulas (\ref{Kubo-GR}) and (\ref{Kubo-rel-2}) with those for the shear viscosity
and the relaxation time so that $\eta$, $\tau_\pi$, and $\kappa$ are eliminated.
As shown in Appendix \ref{app:decomp}, 
the ${\bf k} \to 0$ limit of the pressure-pressure correlation function accomplishes just that,
\ba
\bar{G}_R^{PP}(\omega,0)
& = &
{P\over 3} + \bar G_L(\omega, 0) - {4\over 3}\bar{G}_R^{xy,xy}(\omega, 0)\nn \\
& = &
{P\over 3} + i\omega\zeta - \zeta\tau_{\Pi}\omega^2 +
\mathcal{O}(\omega^3).
\ea
Here the pressure operator is defined as $\hat{P} = \delta_{ij}\hat{T}^{ij}/3$. 
Hence our final Kubo formulas 
for the bulk viscosity and the relaxation time are
\ba
\label{Kubo-final-1}
\zeta 
&=& 
\lim_{\omega \rightarrow 0} \lim_{{\bf k} \rightarrow 0} 
\frac{1}{\omega} \text{Im}\,\bar{G}_R^{PP}(\omega, {\bf k}), \\
\label{Kubo-final-2}
\zeta \tau_\Pi 
&=& 
-\frac{1}{2} \lim_{\omega \rightarrow 0} \lim_{{\bf k} \rightarrow 0} 
\partial^2_\omega \text{Re}\, \bar{G}_R^{PP}(\omega, {\bf k}).
\ea
The formula (\ref{Kubo-final-2}) is especially important, as it consists of
the bulk relaxation time and encodes the prescription on
how to compute it. 

While the Kubo formulas for the linear second-order viscous hydrodynamics have been consistently derived here, it is worth mentioning that the stress-energy correlation functions were also examined in \cite{Young:2013fka}. In the current paper we focus on derivation of the second-order fluid dynamics
from the general analytic properties of the correlation functions while in Ref.~\cite{Young:2013fka} the correlation functions were obtained as the Green functions of the Israel-Stewart type second-order hydrodynamics.

\section{Shear relaxation time in the scalar field theory}
\label{sec-diagram}

We perform here the perturbative analysis of the stress-energy tensor
response functions.
The study is done in the leading order for the
massless real scalar quantum field theory\footnote{
Our analysis can be easily generalized to the massive case by just
substituting $m_{\rm th}^2 \to m_{\rm phys}^2 + m_{\rm th}^2$
where $m_{\rm phys}$ is the physical mass.
}
with the Lagrangian 
\ba
\label{lagrangian-scalar}
\mathcal{L}=
\frac{1}{2}\partial^\mu \phi \partial_\mu \phi - \frac{\lambda}{4!} \phi^4,
\ea
where $\lambda$ is the coupling constant, which is assumed to be small.
Since the scalar field is real, there are no conserved number or charge
operators coupled to the chemical potential. 

In leading order the scalar field dynamics is governed by $2 \leftrightarrow 2$ 
scatterings which give rise only to the shear viscosity
effects being of the order $\mathcal{O}(\lambda^{-2}T^3)$ where $T$ is the
temperature \cite{Jeon:1995zm}. The bulk viscosity strictly vanishes
in the conformal limit. 
Since the conformal symmetry in the scalar theory 
is broken by the nonzero $\beta$ function, the bulk viscosity is much smaller than
the shear viscosity $\zeta = \mathcal{O}( \lambda T^3)$ \cite{Jeon:1995zm}.
It also requires inclusion of number changing inelastic processes at higher orders in
the coupling constant.

Here we work in the leading order of the expansion of the response function
and provide a systematic analysis to compute the shear relaxation
time. To find it we make use of the formula (\ref{Kubo-eta-tau-shear}), where we ignore the
coefficient $\kappa$ since it only
scales as $\mathcal{O}(\lambda^0 T^2)$. By evaluating the real and imaginary parts of 
$\bar G_R^{xy,xy}$ we are able to get $\eta\tau_\pi$ and $\eta$, 
respectively, and then extract the shear relaxation time.
In the forthcoming derivation we employ the closed time path
(Keldysh-Schwinger) formalism which is briefly summarized in
Appendix~\ref{KS-form}. 
The analysis of the four-point functions in Sec. \ref{subsec-ladder}
closely follows \cite{Wang:2002nba}. We also adopt
the notations and sign conventions for the real-time $n$-point functions
from the same reference throughout this section.

\subsection{Definition of the retarded response function}
\label{subsec-def}

Since any response function $\bar G_R$ differs from the
standard retarded Green function due to the Ward identity let us start with
the definition of $\bar G_R^{ij,mn}$. Making allowance for the Ward
identity (\ref{Ward-id-coord}), $\bar G_R^{ij,mn}$ is defined by
\ba
\label{GR-T-xy-W}
\bar G_R^{ij,mn}(x,y) 
&=&
-\delta^{(4)}(x-y)\big(\delta^{jm}\langle \hat T^{in}(y)
\rangle+\delta^{jn}\langle \hat T^{im}(y)\rangle
-\delta^{ij}\langle \hat T^{mn}(y)\rangle  \big) \\ \nn
&&
- i \theta(x_0-y_0) \langle [\hat T^{ij}(x),\hat T^{mn}(y)] \rangle.
\ea
In the equilibrium state, the rotational invariance provides that
$\langle \hat T^{ij} \rangle=\delta^{ij}P$, where $P$ is the thermodynamic
pressure, so that the retarded Green function becomes
\ba
\label{GR-T-xy}
\bar G_R^{ij,mn}(x,y)
&=&
-\delta^{(4)}(x-y)P(y)(\delta^{jm}\delta^{in}+\delta^{jn}\delta^{im}-\delta^{ij}\delta^{mn})
\\ \nn
&&
- i \theta(x_0-y_0) \langle [ \hat T^{ij}(x),\hat T^{mn}(y)] \rangle.
\ea 
Analogously, the advanced Green function is
\ba
\label{GA-T-xy}
\bar G_A^{ij,mn}(x,y)
&=&
-\delta^{(4)}(x-y)P(y)
(\delta^{im}\delta^{jn}+\delta^{in}\delta^{jm}-\delta^{ij}\delta^{mn})\\
\nn
&&
+ i \theta(y_0-x_0) \langle [ \hat T^{ij}(x), \hat T^{mn}(y)]
\rangle.\ea
The commutator in Eqs.~(\ref{GR-T-xy}) and (\ref{GA-T-xy}) consists of the
Wightman functions and should be understood as 
\ba
\label{W-fun}
\langle [ \hat T^{ij}(x),\hat T^{mn}(y)] \rangle 
&=& 
\langle \hat T^{mn}_2(y) \hat T^{ij}_1(x) \rangle - \langle \hat
T^{ij}_2(x) \hat T^{mn}_1(y) \rangle ,
\ea
where the reading from right to left inside the brackets should be
understood as the evolution from the initial to final state; the
indices 1 and 2 locate the operators on the upper (earlier)
or lower (later) branch of the Keldysh time contour, respectively.
The stress tensor operator is defined by
\ba
\label{def-T}
\hat T^{ij} = \hat T^{ij}_{kin} - \frac{1}{3} \delta^{ij} \mathcal{L}
\ea
with the Lagrangian given by (\ref{lagrangian-scalar}). 
Rewriting the Lagrangian as
\be
{\cal L} = 
-{1\over 2}\phi E[\phi] + {\lambda\over 4!}\phi^4,
\ee
where
\ba
E[\phi] = \partial^\mu \partial_\mu \phi + \frac{\lambda}{3!} \phi^3 
\ea
is the equation of motion for the field operator, we see that 
within the thermal average, $\langle {\cal L}\rangle = O(\lambda)$. Hence
the leading 
order stress-energy tensor is dominated by the kinetic term
\ba
\label{field-tensor-sc}
\hat T^{ij}(x) = \partial^i \phi(x) \partial^j \phi(x) +
\mathcal{O}(\lambda).
\ea
Accordingly, the Wightman functions read
\ba
\label{TT-ex}
\langle \hat T_2^{mn}(y) \hat T_1^{ij}(x) \rangle 
&=& 
\langle \partial^n \phi_2(y) \partial^m \phi_2(y) \partial^j 
\phi_1(x)\partial^i \phi_1(x) \rangle, \\
\label{TT-ex}
\langle \hat T_2^{ij}(x) \hat T_1^{mn}(y) \rangle 
&=& 
\langle \partial^j \phi_2(x) \partial^i \phi_2(x) 
\partial^n \phi_1(y)\partial^m \phi_1(y) \rangle,
\ea
which may be expressed in terms of the four-point Green functions as
\ba
\label{TT-ex-4p-gf}
i \langle \hat T_2^{mn}(y) \hat T_1^{ij}(x) \rangle 
&=& 
\partial^i_{x_1} \partial^j_{x_2} \partial^m_{y_1} \partial^n_{y_2} 
G_{1122}(x_1,x_2;y_1,y_2)\big|_{\begin{subarray}{l}x_1=x_2=x \\ 
y_1=y_2=y\end{subarray}}, \\
\label{TT-ex-4p-gh}
i \langle \hat T_2^{ij}(x) \hat T_1^{mn}(y) \rangle 
&=& 
\partial^i_{x_1} \partial^j_{x_2} \partial^m_{y_1} \partial^n_{y_2} 
G_{2211}(x_1,x_2;y_1,y_2)\big|_{\begin{subarray}{l}x_1=x_2=x \\ 
y_1=y_2=y\end{subarray}}.
\ea
The four-point Green functions are defined as
\ba
\label{4p-GF}
i^3 G_{1122}(x_1,x_2;y_1,y_2) 
&=&
\big\langle {\mathcal T}_a \{ \phi_2(y_2) \phi_2(y_1) \} 
\mathcal{T}_c \{ \phi_1(x_2) \phi_1(x_1) \} \big\rangle,  \\
i^3 G_{2211}(x_1,x_2;y_1,y_2) 
&=& 
\big\langle {\mathcal T}_c \{ \phi_1(y_2) \phi_1(y_1) \} 
\mathcal{T}_a \{ \phi_2(x_2) \phi_2(x_1) \} \big\rangle .
\ea
Then the retarded and advanced Green functions 
of the stress-energy tensors
take the forms
\ba
\label{GR-T-xy-ret}
\bar G_R^{ij,mn}(x,y)
&=&
-\delta^{(4)}(x-y)P(y)(\delta^{jm}\delta^{in}+\delta^{jn}\delta^{im}-\delta^{ij}\delta^{mn})
\\ \nn
&&
-\theta(x_0-y_0)
\partial^i_{x_1} \partial^j_{x_2} \partial^m_{y_1}
\partial^n_{y_2} 
\big(G_{1122}(x_1,x_2;y_1,y_2) - G_{2211}(x_1,x_2;y_1,y_2) \big)
\big|_{\begin{subarray}{l}x_1=x_2=x \\ 
y_1=y_2=y\end{subarray}},
\ea
\ba
\label{GA-T-xy-adv}
\bar G_A^{ij,mn}(x,y)
&=&
-\delta^{(4)}(x-y)P(y)(\delta^{im}\delta^{jn}+\delta^{in}\delta^{jm}-\delta^{ij}\delta^{mn})\\
\nn
&&
+\theta(y_0-x_0) \partial^i_{x_1} \partial^j_{x_2} \partial^m_{y_1}
\partial^n_{y_2} 
\big(G_{1122}(x_1,x_2;y_1,y_2) - G_{2211}(x_1,x_2;y_1,y_2) \big)
\big|_{\begin{subarray}{l}x_1=x_2=x \\ 
y_1=y_2=y\end{subarray}}
\ea
where the limits $x_1\to x_2$ and $y_1\to y_2$ must be taken
after the derivatives.

\subsection{Real part of the retarded Green function in the free field theory}
\label{subsec-one-loop}

To find the real part of $\bar G^{ij,mn}$ we continue our
considerations to study the one-loop diagram which appears in the free scalar quantum field 
theory; this is done mostly to work out the details. 
\begin{figure}[!h]
\centering
\includegraphics*[width=0.3\textwidth]{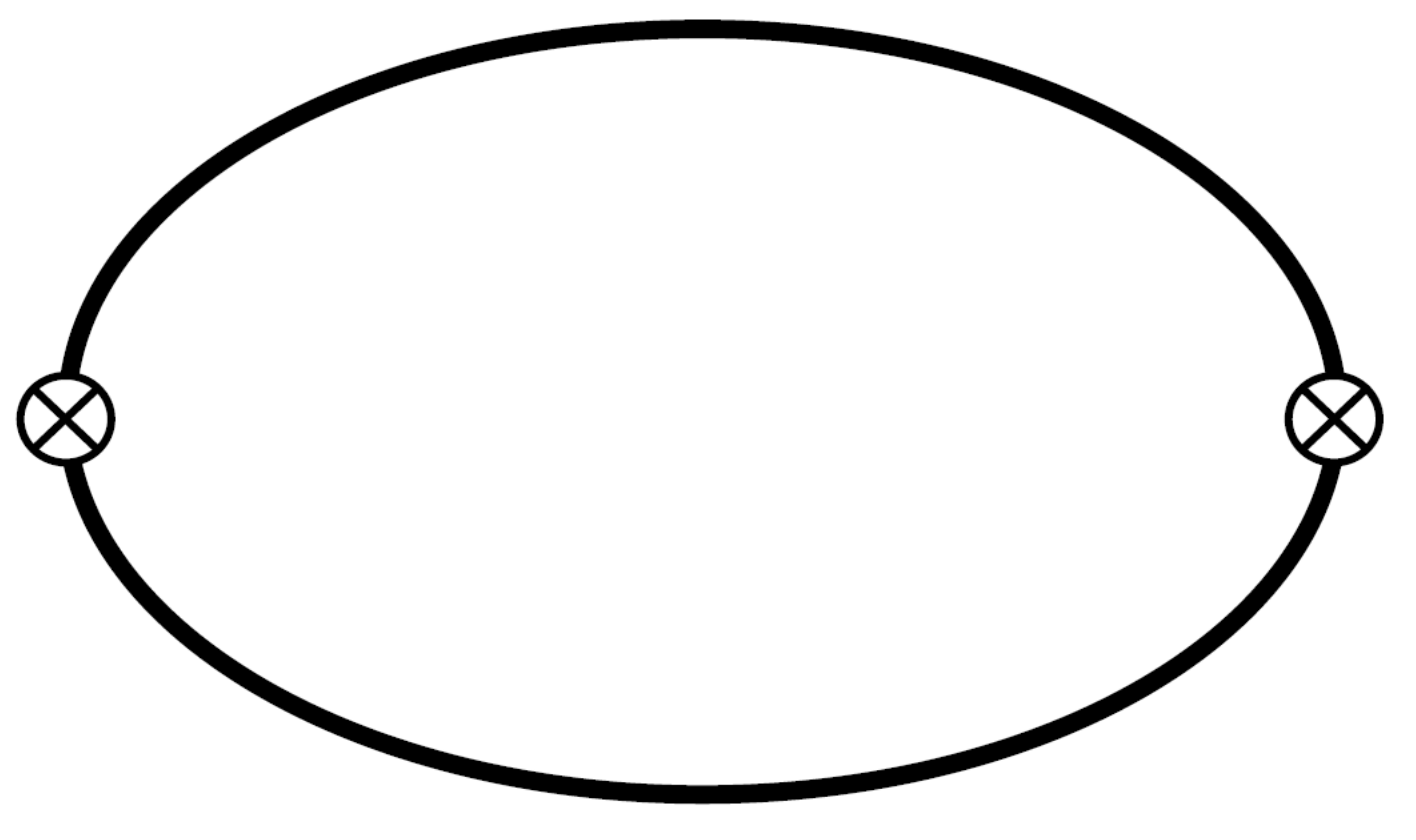}
\caption{One-loop diagram of the free scalar field theory representing the contribution to 
the retarded Green function of the energy-momentum tensor operator.}
\label{fig-diagram}
\end{figure}
This diagram is shown in Fig.~\ref{fig-diagram} and 
it comes from the 
disconnected parts of four-point
Green functions which are nothing but the products of the Wightman functions,
\ba
i^3 G_{1122}(x_1,x_2;y_1,y_2) 
&=& 
\big\langle \phi_2(y_1) \phi_1(x_2) \big\rangle 
\big\langle \phi_2(y_2) \phi_1(x_1) \big\rangle +
\big\langle \phi_2(y_1) \phi_1(x_1) \big\rangle 
\big\langle \phi_2(y_2) \phi_1(x_2) \big\rangle, \;\;\;\\
i^3 G_{2211}(x_1,x_2;y_1,y_2) 
&=& 
\big\langle \phi_2(x_2) \phi_1(y_1) \big\rangle 
\big\langle \phi_2(x_1) \phi_1(y_2) \big\rangle +
\big\langle \phi_2(x_1) \phi_1(y_1) \big\rangle 
\big\langle \phi_2(x_2) \phi_1(y_2) \big\rangle .\;\;\;
\ea
When the points are joined so that $x_1=x_2=x$ and $y_1=y_2=y$, we get
\ba
i^3 G_{1122}(x,x;y,y) 
&=& 
2i\Delta_{12}(x,y) i\Delta_{21}(y,x), \qquad \\ 
i^3 G_{2211}(x,x;y,y) 
&=& 
2 i\Delta_{21}(x,y)i\Delta_{12}(y,x),
\ea
where we defined $\Delta_{12}(x,y) = \langle\phi_1(x)\phi_2(y)\rangle$ 
and $\Delta_{21}(x,y) = \langle\phi_2(x)\phi_1(y)\rangle$
also used the fact that $\Delta_{12}(x,y)=\Delta_{21}(y,x)$.
Since the system considered is
homogeneous, the two-point functions depend on $x$ and $y$ only through
their difference $x-y$. Putting
all these facts together, we get the
retarded function $\bar G_R^{ij,mn}$ in the form
\ba
\label{GR-T-xy-delta}
\bar G_R^{ij,mn}(x-y)
= 
-\delta^{(4)}(x-y)P(\delta^{im}\delta^{jn}+\delta^{in}\delta^{jm}-\delta^{ij}\delta^{mn})
\qquad\qquad\qquad\qquad\qquad\qquad \\ \nn 
+2i\theta(x_0-y_0) 
\big(\partial_x^j \partial_y^m \Delta_{12}(x-y) \partial_x^i
\partial_y^n
\Delta_{21}(y-x)
- \partial_x^j \partial_y^m \Delta_{21}(x-y) \partial_x^i
\partial_y^n\Delta_{12}(y-x) \big).
\ea

At this point, it is more convenient to change the basis from $\phi_{1,2}$ to
$\phi_{r,a}$ defined by
\ba
\phi_r(x) &=& {\phi_1(x) + \phi_2(x)\over 2},
\\
\phi_a(x) &=& \phi_1(x) - \phi_2(x).
\ea
Using the relations between the Green functions in the (1,2) and $(r,a)$
bases expressed by (\ref{rr-12})--(\ref{aa-12}) and (\ref{11-ra})--(\ref{22-ra}), we find 
\ba
\label{GR-T-xy-delta-kk}
\bar G_R^{ij,mn}(x-y)
=
-\delta^{(4)}(x-y)P(\delta^{im}\delta^{jn}+\delta^{in}\delta^{jm}-\delta^{ij}\delta^{mn})
\qquad\qquad\qquad\qquad\qquad \\ \nn 
-i \big(\partial_x^j \partial_y^m \Delta_{ra}(x-y) \partial_x^i
\partial_y^n \Delta_{rr}(y-x)
+ \partial_x^j \partial_y^m \Delta_{rr}(x-y) \partial_x^i
\partial_y^n\Delta_{ar}(y-x) \big).
\ea
Performing the Fourier transform we next get
\ba
\label{GR-T-fou-2}
\bar G^{ij,mn}_R(k)  
&=& 
-P(\delta^{im}\delta^{jn}+\delta^{in}\delta^{jm}-\delta^{ij}\delta^{mn}) \\ \nn
&& 
-i \int \frac{d^4 p}{(2\pi)^4} p^i p^n (p+k)^j (p+k)^m
\Big[ \Delta_{rr}(p)\Delta_{ra}(p+k) + \Delta_{rr}(p+k) \Delta_{ar}(p)\Big],
\ea
where $k\equiv(k_0,{\bf k})\equiv(\omega,{\bf k})$.
And analogously
$\bar G^{ij,mn}_A(k)$ is
\ba
\label{GA-T-fou-2}
\bar G^{ij,mn}_A(k)  
&=& 
-P(\delta^{im}\delta^{jn}+\delta^{in}\delta^{jm}-\delta^{ij}\delta^{mn}) \\ \nn
&& 
-i \int \frac{d^4 p}{(2\pi)^4} p^i p^n (p+k)^j (p+k)^m
\Big[ \Delta_{rr}(p)\Delta_{ar}(p+k) + \Delta_{rr}(p+k) \Delta_{ra}(p)\Big].
\ea

The functions $\Delta_{ra}(p)$ and $\Delta_{ar}(p)$ are the usual
retarded and advanced two-point Green functions which are of the following forms:
\ba
\label{ret-adv-fu1}
\Delta_{ra}(p)&=&\frac{1}{(p_0+i\epsilon)^2 - {\bf p}^2},\\
\Delta_{ar}(p)&=&\frac{1}{(p_0-i\epsilon)^2 - {\bf p}^2},
\ea
and they satisfy
\ba
\label{ret-adv-rell}
\Delta_{ra}(p)=\Delta_{ar}^*(p).
\ea
$\Delta_{rr}(p)$ is the autocorrelation function and it is the only
function where a distribution function explicitly enters. In thermal equilibrium all
three of these functions are related via the fluctuation-dissipation theorem,
\ba
\label{fl-dis-theorem}
\Delta_{rr}(p)=N(p^0)[\Delta_{ra}(p)-\Delta_{ar}(p)],
\ea
where $N(p^0)=1+2n(p_0)$ and $n(p^0)=1/(e^{\beta p^0}-1)$ is the Bose
distribution function with $\beta$ being the inverse of temperature $T$.

The functions $\bar G^{xy,xy}_R(k)$ and $\bar G^{xy,xy}_A(k)$ are then 
obtained by setting $i=m=x$ and $j=n=y$ in Eqs.~(\ref{GR-T-fou-2})
and (\ref{GA-T-fou-2}). We also choose ${\bf k}=(0,k_y,0)$
to use our analysis in Sec. \ref{subsec-transverse}.
The real
and imaginary parts of $\bar G^{xy,xy}_R(k)$ are then obtained by
the sum and the difference of $\bar G^{xy,xy}_R(k)$ and $\bar G^{xy,xy}_A(k)$, 
respectively. In the vanishing momentum limit they
are as follows:
\ba
\label{GR-GA-r}
&&\lim_{k_y \rightarrow 0} \textrm{Re } \bar
G^{xy,xy}_R(\omega,k_y)+P=
-\frac{i}{2} \int \frac{d^4 p}{(2\pi)^4}
p_x^2 p_y^2
\qquad\qquad \\ \nn
&& 
\qquad\qquad\qquad 
\times \Big[ \Delta_{rr}(p)[\Delta_{ra}(p+k) + \Delta_{ar}(p+k) )]
+ \Delta_{rr}(p+k)[\Delta_{ar}(p) + \Delta_{ra}(p)] \Big], \\
\label{GR-GA-i}
&& 
\lim_{k_y \rightarrow 0} \textrm{Im }\bar G^{xy,xy}_R(\omega,k_y)
= -\frac{1}{2} \int \frac{d^4 p}{(2\pi)^4} p_x^2 p_y^2
\\ \nn
&&
\qquad\qquad\qquad 
\times \Big[ \Delta_{rr}(p)[\Delta_{ra}(p+k) - \Delta_{ar}(p+k) ]
+ \Delta_{rr}(p+k)[\Delta_{ar}(p) - \Delta_{ra}(p)] \Big],
\ea
where in the right hand side $k = (\omega, \mathbf{0})$.

\subsection{Pinching poles}
\label{subsec-pinching}

Provided that the fluctuation-dissipation theorem of the form
(\ref{fl-dis-theorem}) is applied to the formulas (\ref{GR-GA-r}) and
(\ref{GR-GA-i}), there appear, in particular, terms of products of propagators with four
poles lying symmetrically on both sides of the real axis in the complex
$p_0$ plane if the small $\omega$ limit is used. These
poles give rise to the pinching of the integration contour by the
poles lying on opposite sides of the real energy axis, that is, the pinching pole
effect. These terms indeed may be evaluated as
\ba
\label{singularity}
\int \frac{dp_0}{2\pi}\Delta_{ra}(p)\Delta_{ar}(p) 
&\sim& 
\int \frac{dp_0}{2\pi} \frac{1}{(p_0+i\epsilon)-|{\bf p}|}\;
\frac{1}{(p_0-i\epsilon)-|{\bf p}|}  \sim \frac{1}{\epsilon},
\ea
so that they produce a singularity as $\epsilon \rightarrow 0$. In a
non-interacting theory such an effect is natural and it means that
since the emerged excitation is not subject to collisions it can propagate
indefinitely long. It is reflected by $\delta$ functions carried by the spectral
density. The width of such a peak, which is inversely proportional to the
lifetime of the excitation, is vanishingly small. 
This implies that
in the free theory there is no transport of
conserved quantities and consequently transport coefficients cannot be defined. 
In an interacting system, transport
coefficients are finite due to 
the finite mean free path (or lifetime) of a 
propagating excitation until it suffers from scatterings with constituents of 
the thermal bath. Thus, in thermal weakly interacting medium the spectral density
can be approximated by Lorentzians
\cite{Jeon:1994if}
\ba
\label{spectral-lorentzian}
\rho(p)=\frac{1}{2E_p} \Big(\frac{2\Gamma_p}{(p_0-E_p)^2 +
\Gamma_p^2}- \frac{2\Gamma_p}{(p_0+E_p)^2 + \Gamma_p^2} \Big),
\ea
where $E_p$ is 
the quasi-particle excitation energy
and $\Gamma_p$ is
the thermal width. The origin of such a form of the spectral density may
be also understood if one uses the resummed
propagators. These propagators carry information on the interaction of a given
particle with the medium in terms of
the self-energy $\Sigma=\textrm{Re}\,\Sigma
+i \textrm{Im}\,\Sigma$. 
They are defined as
$\Delta_{ra}(p) = [p^2 - m^2 - \Sigma(p)]^{-1}$
and $\Delta_{ar}(p)=\Delta^*_{ra}(p)$, where $m$ is
the mass. 

In this paper,
we study massless theory, but the real part of the self-energy in the lowest order
does not vanish. The leading order diagram
is the tadpole diagram, which is momentum independent, 
and may be identified 
as the thermal mass ($m_{\rm th}$) squared. The spectral density is then given in
terms of the resummed retarded and advanced propagators as
\ba
\label{spectral-den-prop}
\rho(p)= i[\Delta_{ra}(p) - \Delta_{ar}(p)].
\ea
When the spectral density has sharp peaks near
$p_0=\pm E_p$, we can then say that
the dispersion relation is $E_p^2={\bf p}^2 + m_{\rm th}^2$, and the thermal width
is related to the imaginary part of the
self-energy as $\Gamma_p=\frac{\textrm{Im}\,\Sigma(E_p, |{\bf p}|)}{2E_p}$.
Consequently, the resummed
retarded and advanced propagators can be approximated as
\ba
\label{propagators-dressed-ra}
\Delta_{ra}(p) &=& \frac{1}{(p_0+i\Gamma_p)^2 - E_p^2}, \\
\label{propagators-dressed-ar}
\Delta_{ar}(p) &=& \frac{1}{(p_0-i\Gamma_p)^2 - E_p^2}.
\ea
Appearance of $\Gamma_p$ in the propagators shifts the pinching poles
away from the real axis in the complex $p_0$ plane, which regulates the
singularities in (\ref{GR-GA-r}) and (\ref{GR-GA-i}) making the
integral finite. The pinching pole contribution is then of the order $\mathcal{O}(1/\Gamma_p)$. What is more, the terms of the type $\Delta_{ra}(p)\Delta_{ra}(p)$ and
$\Delta_{ar}(p)\Delta_{ar}(p)$ have poles on the same side of the
real energy axis and thus they give much smaller contribution to the
expressions (\ref{GR-GA-r}) and (\ref{GR-GA-i}) than the pinching poles, and may
be safely ignored in further computations. The omission of these terms 
constitutes the pinching pole approximation.

\begin{figure}[!h]
\centering
\includegraphics*[width=0.8\textwidth]{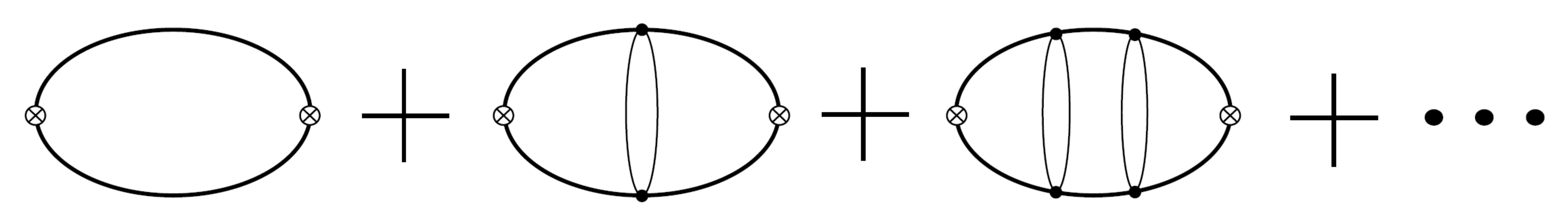}
\caption{Resummation of ladder diagrams. The insertions of the energy-momentum 
tensor operator $\hat T^{xy}$ is denoted by the crossed dots and black dots are the vertices 
with the coupling constant $\lambda$.}
\label{fig-GF1}
\end{figure}

Replacement of bare propagators by dressed ones means that we need to deal with
the skeleton expansion where propagators are dressed and
vertices remain bare. Here, we are to study the first loop of this expansion.
However, since the thermal width is related to the imaginary part of a
self-energy, some complications arise. In the weakly coupled 
$\lambda \phi^4$ theory the lowest contribution
to $\textrm{Im}\,\Sigma$ comes from a two-loop diagram which is of the
order $\mathcal{O}(\lambda^2)$ and since the pinching pole contribution dominates,
the one-loop diagram is of the order
$\mathcal{O}(1/\lambda^2)$ \cite{Jeon:1992kk}. However, one realizes 
that there may be momentum exchange
between the side rails of the loop. This is represented by the one-loop rungs
connecting the two side rails 
as shown in Fig.~\ref{fig-GF1}.

Each rung introduces a factor of 
$\lambda^2$ coming from the vertices and a factor of the order
$\mathcal{O}(1/\lambda^2)$ coming from the pinching poles introduced by
the additional pair of propagators. 
Therefore, all such multi-loop 
ladder diagrams contribute at the leading order.
They must be resummed to give the full result in the leading order. 

The situation described above holds when the single transport coefficient,
such as the shear viscosity, is analyzed. In case of the combination $\eta\tau_\pi$,
it gets more involved and it will be discussed in the next part of this work.

\subsection{Evaluation of $\eta$ and $\eta\tau_\pi$ in the one-loop limit}
\label{subsec-one-loop}

Before we include all ladder diagrams let us consider first only the one-loop diagram
with the resummed propagators. This is illuminating as we can find the
typical scales of $\eta$ and $\eta\tau_\pi$.

The shear viscosity is related to the imaginary part of
the relevant retarded Green function.
It can be calculated in a few ways which
are related to different choices of 
the correlation function. Usually, it
is examined from the imaginary part of the Green function of the
traceless, spatial part of the stress-energy tensor $\pi^{ij}$. It may be also
computed in terms of the stress-stress function $\bar G_R^{xy,xy}$ as
shown by the Kubo formula (\ref{KF-1r}). We employ here
the latter choice. Applying the fluctuation-dissipation theorem~(\ref{fl-dis-theorem})
and the pinching pole approximation to
Eq.~(\ref{GR-GA-i}), the imaginary part is then given by
\ba
\label{im-sm}
\lim_{k_y \rightarrow 0} 
\textrm{Im} \bar G^{xy,xy}_R(\omega, k_y)
&=&  
\frac{1}{2} \int \frac{d^4p}{(2\pi)^4}  p^2_x p^2_y \\ \nn
&& 
\times [N_p-N_{p+k}] \Big( \Delta_{ar}(p+k) \Delta_{ra}(p) +
\Delta_{ra}(p+k)
\Delta_{ar}(p)\Big)
\ea
with the propagators $\Delta_{ra}$ and $\Delta_{ar}$ being dressed
and given by (\ref{propagators-dressed-ra}) and (\ref{propagators-dressed-ar})
and $k = (\omega, \mathbf{0})$ in the right hand side.
The real part of the retarded Green function (\ref{GR-GA-r}) is
\ba
\label{real-sm}
&&
\lim_{k_y \rightarrow 0} \textrm{Re } \bar G^{xy,xy}_R(\omega,k_y) 
+P=
-\frac{i}{2} \int \frac{d^4 p}{(2\pi)^4} p_x^2 p_y^2
\qquad\qquad \\ \nn
&& 
\qquad\qquad \times [N_p-N_{p+k}]
\Big( \Delta_{ar}(p+k) \Delta_{ra}(p) - \Delta_{ra}(p+k)
\Delta_{ar}(p)\Big).
\ea

In pursuit of $\eta=\frac{1}{\omega}\text{Im} \bar G^{xy,xy}_R(\omega,k_y)|_{\omega,k_y \to 0} 
= \partial_\omega \text{Im} \bar G^{xy,xy}_R(\omega,k_y)|_{\omega,k_y \to 0}$,
we find
\ba
\label{po}
\lim_{\omega,k_y \to 0} \partial_\omega \text{Im} \bar G^{xy,xy}_R(\omega,k_y)=
\lim_{\omega,k_y \to 0}\int \frac{d^4p}{(2\pi)^4}
p^2_x p^2_y \,\partial_\omega [N_p-N_{p+k}]
 \Delta_{ar}(p) \Delta_{ra}(p).
\ea
Realizing that
\ba
\label{N}
\lim_{\omega \to 0} \partial_\omega [N_p-N_{p+k}] = 2\beta n(p_0)(n(p_0)+1),
\ea
the one-loop shear viscosity is
\ba
\label{eta-kk}
\eta_{\rm 1-loop}
&=& 2 \beta \int \frac{d^4p}{(2\pi)^4} \;
p^2_x p^2_y \, n(p_0)(n(p_0)+1) 
\Delta_{ra}(p) \Delta_{ar}(p).
\ea

The action of the second-order derivative and inclusion of the factor
$-1/2$ to the real part given by the formula (\ref{real-sm}), as dictated by the Kubo
formula (\ref{Kubo-eta-tau-shear}), leads us to the following equation:
\ba
\label{eta-tau-kk}
\left.\eta\tau_\pi\right|_{\rm 1-loop}
&=&  
-\frac{i}{2}\lim_{\omega,k_y \to 0}\int \frac{d^4p}{(2\pi)^4}
p^2_x p^2_y \,\partial_\omega [N_p-N_{p+k}]
\partial_\omega \Big( \Delta_{ar}(p+k) \Delta_{ra}(p) -
\Delta_{ra}(p+k), \Delta_{ar}(p)\Big). \nn \\
\ea
Taking Eq. (\ref{N}) into account and
\ba
\label{der-ra-ar}
\lim_{\omega,k_y \to 0}\partial_\omega \Big( \Delta_{ar}(p+k) \Delta_{ra}(p) -
\Delta_{ra}(p+k), \Delta_{ar}(p)\Big)=4i\Gamma_p(p^2_0+\Gamma^2_p+E_p^2) \Delta^2_{ra}(p) \Delta^2_{ar}(p),\;\;\;
\ea
the expression (\ref{eta-tau-kk}) becomes
\ba
\label{eta-tau-kk-1}
\left.\eta\tau_\pi\right|_{\rm 1-loop}
 &=&  4\beta \int \frac{d^4p}{(2\pi)^4}
p^2_x p^2_y \; n(p_0)(n(p_0)+1) \;
\Gamma_p(p^2_0+\Gamma^2_p+E_p^2) \Delta^2_{ra}(p) \Delta^2_{ar}(p).
\ea

The frequency integrals to perform in (\ref{eta-kk}) and (\ref{eta-tau-kk-1}) are
\ba
\label{eta-kk-2}
I_1 = 
\int \frac{dp_0}{2\pi} n(p_0)(n(p_0)+1) \;
\frac{1}{[(p_0+i\Gamma_p)^2-E_p^2][(p_0-i\Gamma_p)^2-E_p^2]},
\qquad
\ea
\ba
\label{eta-tau-kk-2}
I_2 =
\int \frac{dp_0}{2\pi} n(p_0)(n(p_0)+1) \;
\frac{\Gamma_p(p^2_0+\Gamma^2_p+E_p^2)}{[(p_0+i\Gamma_p)^2-E_p^2]^2[(p_0-i\Gamma_p)^2-E_p^2]^2}.
\qquad
\ea
The integrands in Eqs.~(\ref{eta-kk-2}) and (\ref{eta-tau-kk-2}) have four poles at 
$p_1= i\Gamma_p +E_p$, $p_2= i\Gamma_p -E_p$, $p_3=- i\Gamma_p +E_p$,
 and $p_4=- i\Gamma_p -E_p$. In Eq.~(\ref{eta-kk-2}) the poles are simple poles 
while 
in Eq.~(\ref{eta-tau-kk-2}) they are double poles. 
By using the residue theorem and closing the contour in the upper-half plane, the sum of
the residua in Eq.~(\ref{eta-kk-2}) is found in the leading order of $\Gamma_p/E_p$ as
\ba
\label{grouped-terms-eta}
I_1 = \frac{n(E_p)(n(E_p)+1)}{4E_p^2 \Gamma_p }.
\ea
In Eq.~(\ref{eta-tau-kk-2}) we handle the second-order poles. We recall
that the residua of a function with second-order poles contain the
derivative with respect to the complex argument. 
Thus, when the contour integration is carried out,
the expression (\ref{eta-tau-kk-2}) becomes
\ba
\label{grouped-terms}
I_2 = \frac{n(E_p)(n(E_p)+1)}{16 E_p^2 \Gamma_p^2 }.
\ea
Finally, the formula~(\ref{eta-kk}) for the shear viscosity is
\ba
\label{eta-fullBS-final}
\eta_{\rm 1-loop} &=&  
\frac{\beta}{2} \int \frac{d^3p}{(2\pi)^3} p_x^2 p_y^2
\frac{n(E_p)(n(E_p)+1)}{E_p^2
\Gamma_p }
\ea 
and the product of the shear viscosity and its relaxation time is
\ba
\label{eta-tau-fullBS-final}
\left.\eta\tau_\pi \right|_{\rm 1-loop}
&=& \frac{\beta}{4} \int \frac{d^3p}{(2\pi)^3} p_x^2
p_y^2
\frac{n(E_p)(n(E_p)+1)}{E_p^2 \Gamma_p^2 }.
\ea
At this level one immediately notices that the shear relaxation time
scales as $1/\Gamma_p$. The computation of its value is given in
Sec.~\ref{sec-discussion}.

\subsection{Summation over multiloop diagrams}
\label{subsec-ladder}

The one-loop limit is, however, not sufficient and we need to resum ladder diagrams 
which requires us to manipulate the connected 4-point Green functions
as well. To do so we employ the definitions (\ref{ra-12}) and
(\ref{ar-12}) to get the retarded and advanced four-point Green functions as
\ba
\label{GR-4}
\tilde G_R(x_1,x_2,x_3,x_4) =
G_{1111}(x_1,x_2,x_3,x_4)-G_{1122}(x_1,x_2,x_3,x_4),
\ea
\ba
\label{GA-4}
\tilde G_A(x_1,x_2,x_3,x_4) =
G_{1111}(x_1,x_2,x_3,x_4)-G_{2211}(x_1,x_2,x_3,x_4).
\ea
The subscripts $R$ and $A$ in $\tilde G_R$ and $\tilde G_A$ 
do not mean that they themselves are four-point retarded and advanced functions.
The subscripts just indicates that 
these functions will become the two-point retarded and advanced 
functions when $x_1$ is identified with $x_2$ and $y_1$ is identified with $y_2$.
The real part is 
\ba
\label{GR-4-re}
\textrm{Re} \; \tilde G_R
&=&
\frac{1}{2}(2G_{1111}-G_{2211}-G_{1122})
= \frac{1}{2}(G_{1111} - G_{2222}),
\ea
where we have used the relation (\ref{rel-12}), and the imaginary part
is
\ba
\label{GR-4-re}
\textrm{Im} \; \tilde G_R
&=&
\frac{1}{2}(G_{2211}-G_{1122}).
\ea
The four-point functions in $(1,2)$ basis may be transformed to the
$(r,a)$ basis using the relation (\ref{invert-rel}). Then, one finds
\ba
\label{GR-4-re-ar}
\textrm{Re}\; \tilde G_R
&=&
\frac{1}{8}(G_{rrra}+G_{rrar}+G_{rarr}+G_{arrr}
+G_{aaar}+G_{aara}+G_{araa}+G_{raaa}), \\
\label{GR-4-im-ar}
\textrm{Im} \; \tilde G_R
&=&
\frac{1}{8}(G_{rrra}+G_{rrar}-G_{rarr}-G_{arrr}
+G_{aaar}+G_{aara}-G_{araa}-G_{raaa}).
\ea
Despite the fact that the shear viscosity, related to the imaginary
part of a Green function, has been studied in literature many times, it may
be illuminating to see some analogies and differences between the real
and imaginary parts. Thus, when deriving the real part of the Green
function we will be referring to $\textrm{Im}\tilde G_R$ quite frequently, as
well. In particular, we will be quoting the results from
\cite{Wang:2002nba}, where $\eta$ was derived in the real-time formalism. 

\begin{figure}[!h]
\centering
\includegraphics*[width=0.8\textwidth]{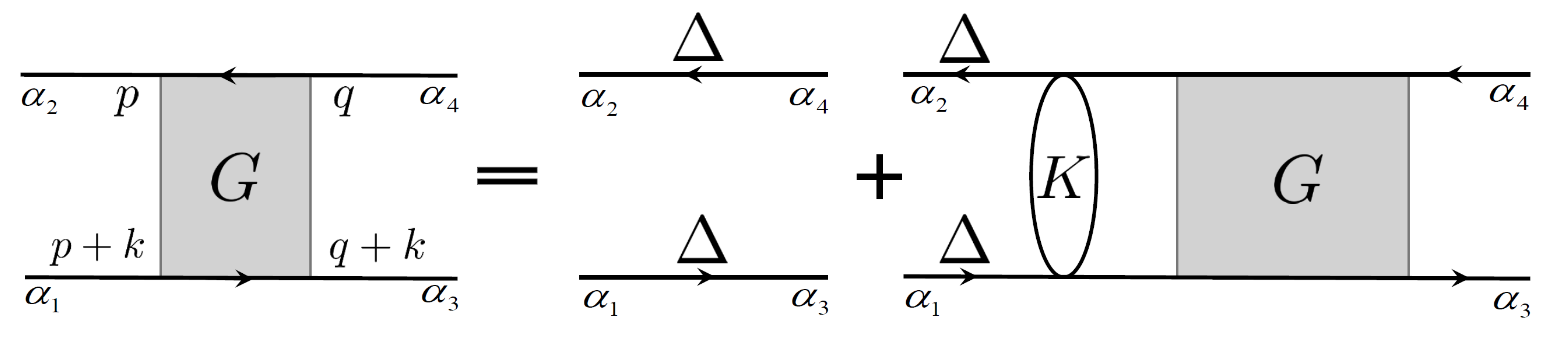}
\caption{Four-point Green function. }
\label{fig-GF}
\end{figure}

Each of the four-point Green functions in Eqs. (\ref{GR-4-re-ar}) and
(\ref{GR-4-im-ar}) couples to the remaining ones. This is shown in
Fig.~\ref{fig-GF} where the disconnected parts represent the one-loop diagram, the rung
is a kernel and the shaded box is a sum of all possible combinations
of four-point Green functions in the ($r,a$) basis. Both the kernel and the
shaded box contribute to an effective vertex. Therefore, Fig.~\ref{fig-GF} presents 
an infinite series of diagrams which can be written as the Bethe-Salpeter 
equation (BSE). For an arbitrary function $G_{\alpha_1\alpha_2\alpha_3\alpha_4}$, the BSE reads
\ba
\label{BSE}
&&
i^3 G_{\alpha_1\alpha_2\alpha_3\alpha_4}(p+k,-p,-q-k,q) \\ \nn
&& \qquad\qquad
= i\Delta_{\alpha_1\alpha_3}(p+k)\,
i\Delta_{\alpha_2\alpha_4}(-p) (2\pi)^4 \delta^4(p-q)
+ i\Delta_{\alpha_1\beta_1}(p+k) i\Delta_{\alpha_2\gamma_1}(-p) \\
\nn&&\qquad\qquad
\times \int \frac{d^4l}{(2\pi)^4} 
K_{\beta_1\gamma_1\beta_4\gamma_4}(p+k,-p,-l-k,l) 
i^3 G_{\beta_4\gamma_4\alpha_3\alpha_4}(l+k,-l,-q-k,q).
\ea
The analytic solution of the BSE is, in general, not readily available.
However, as already mentioned, the use of the Keldysh basis
accompanied by the pinching pole approximation makes it much simpler.

In \cite{Wang:2002nba} it has been shown that by means of the
fluctuation-dissipation theorems (FDT), developed 
in~\cite{Wang:1998wg,Carrington:2006xj}, one may show that the only
contribution to $\textrm{Im} \tilde G_R$ comes from
$\textrm{Im} G_{aarr}$. Even without referring to FDT one may quickly observe that
the functions $G_{aaar}$, $G_{aara}$, $G_{araa}$, and $G_{raaa}$ in
Eqs.~(\ref{GR-4-re-ar}) and (\ref{GR-4-im-ar}) 
do not contribute because these functions must contain at least one $\Delta_{aa}$.

In this way we are left only with $G_{rrra}$, $G_{rrar}$, $G_{rarr}$,
and $G_{arrr}$. As shown in Appendix~\ref{notes}, these functions are
related to $G_{aarr}$ and $G_{rraa}$ by
\ba
\label{combination_1}
G_{arrr}+G_{rarr}&=& [N_p-N_{p+k}]G_{aarr}, \\
\label{combination_2}
G_{rrar}+G_{rrra}&=& [-N_q+N_{q+k}]G_{rraa},
\ea
where $N_p=N(p_0)$. The functions $G_{aarr}$ and $G_{rraa}$ are also related
through FDT.  The relevant relation is \cite{Carrington:2006xj}
\ba
\label{fdt}
(-N_q+N_{q+k})\big(G_{rraa} +N_{p+k}G_{araa}-N_p G_{raaa} \big)
\qquad\qquad\qquad\qquad \\ \nn
= 
(N_p-N_{p+k})\big(G^*_{aarr}- N_{q+k}G^*_{aaar}+N_q G^*_{aara} \big).
\ea
So, ignoring again the contributions from the functions with three
$a$ indices, we obtain $\textrm{Re} \tilde G_R$ and $\textrm{Im}
\tilde G_R$ as follows:
\ba
\label{GR-real}
\textrm{Re} \; \tilde G_R &=& 
\frac{1}{8}[N_p-N_{p+k}] \big(G_{aarr}+G^*_{aarr}\big), \\
\label{GR-imag}
\textrm{Im} \; \tilde G_R &=& 
\frac{1}{8}[N_p-N_{p+k}] \big(G_{aarr}-G^*_{aarr}\big).
\ea

\subsection{Bethe-Salpeter equation for $G_{aarr}$ and $G^*_{aarr}$}
\label{subsec-BSE}

The Bethe-Salpeter equation for $G_{aarr}$ is analyzed in
detail in \cite{Wang:2002nba}. 
Nonetheless, we repeat here the main steps because unlike the
viscosity calculations, the $\eta\tau_\pi$ calculation requires
that the external frequency $k_0=\omega$ be kept until the 
derivatives are taken. 
Fully written out, 
the Bethe-Salpeter equation for $G_{aarr}$ is
\ba
\label{BSE-1}
&&
G_{aarr}(p+k,-p,-q-k,q)= 
-\Delta_{ar}(p+k)\, \Delta_{ra}(p) \Big[ i(2\pi)^4 \delta^4(p-q) \\ \nn
&& \qquad\qquad
+ \int \frac{d^4l}{(2\pi)^4} 
K_{rr\beta_4\gamma_4}(p+k,-p,-l-k,l)
G_{\beta_4\gamma_4rr}(l+k,-l,-q-k,q)
\Big].
\ea
The external momentum $k$ can flow in an arbitrary way along a diagram, that is, it
can enter the rungs but it does not have to. Since all opportunities are
equally possible and lead to the same final result, we have a freedom
to choose what is convenient for computations. Therefore, in
this analysis the external momentum is flowing along the external lower
side rail only, as indicated in
Fig.~\ref{fig-GF}. Then, the kernel couples to
four-point Green functions as
\ba
\label{kernel}
K_{rr\beta_4\gamma_4} G_{\beta_4\gamma_4rr}
=
K_{rraa}G_{aarr}+K_{rrra}G_{rarr}+K_{rrar}G_{arrr},
\ea
where we have taken into account that $K_{rrrr}=0$, which is the
analog of $G_{aaaa}=0$ amputated of the external legs. By truncating 
external legs from $G_{rarr}$ and $G_{arrr}$, that is, by using
the formulas (\ref{Grarr}) and (\ref{Garrr}), 
we find that the expression (\ref{kernel})
becomes
\ba
\label{kernel-2}
K_{rraa}G_{aarr}+K_{rrra}G_{rarr}+K_{rrar}G_{arrr} 
=
(K_{rraa} - N_{l+k}K_{rrra} + N_l K_{rrar})G_{aarr}.
\ea
The Bethe-Salpeter equation for $G_{aarr}$ becomes now
\ba
\label{BSE-2}
&&
G_{aarr}(p+k,-p,-q-k,q)
= -\Delta_{ar}(p+k)\, \Delta_{ra}(p) \Big[ i(2\pi)^4 \delta^4(p-q) \\
\nn
&&\qquad\qquad
+ \int \frac{d^4l}{(2\pi)^4} 
(K_{rraa} -N_{l+k}K_{rrra} +N_l K_{rrar})(p+k,-p,-l-k,l) \\ \nn
&&
\qquad\qquad\qquad \times G_{aarr}(l+k,-l,-q-k,q) \Big].
\ea
To find $G^*_{aarr}$ one just makes a complex conjugate of the
formula (\ref{BSE-2}). Taking into account the analysis of the complex
conjugate procedure of the kernel rungs, shown in Appendix~\ref{notes}, one gets
\ba
\label{BSE-22}
&&G^*_{aarr}(p+k,-p,-q-k,q)= -\Delta_{ra}(p+k)\, \Delta_{ar}(p)
\Big[-i(2\pi)^4 \delta^4(p-q) \\ \nn
&& \qquad
+ \int \frac{d^4l}{(2\pi)^4} 
(K_{rraa} +N_{l+k}K_{rrar} -N_l K_{rrra})(p+k,-p,-l-k,l) \\ \nn
&& \qquad\qquad
\times G^*_{aarr}(l+k,-l,-q-k,q) \Big].
\ea
Note that the combinations of the kernel functions in (\ref{BSE-2}) and (\ref{BSE-22})
become identical in the $k \to 0$ limit.

\subsection{Evaluation of $\eta$ and $\eta\tau_\pi$}
\label{subsec-relax-time}

To calculate the shear viscosity and the shear relaxation time, one needs
to evaluate Eq. (\ref{GR-imag}) with the appropriate derivatives
\ba
\label{GR-4-im-gen}
\lim_{k_y \rightarrow 0} 
\textrm{Im} \bar G^{xy,xy}_R(\omega, k_y)
&=&  
\frac{i}{2} \int \frac{d^4p}{(2\pi)^4}  p_x p_y \int
\frac{d^4q}{(2\pi)^4}  q_x q_y \\ \nn
&& 
\times [N_p-N_{p+k}] \big(G_{aarr} - G^*_{aarr}\big) (p+k,-p,-q-k,q)
\ea
and Eq. (\ref{GR-real}),
\ba
\label{GR-4-re-gen}
\lim_{k_y \rightarrow 0} 
\textrm{Re} \bar G^{xy,xy}_R(\omega,k_y) +P
&=&
\frac{1}{2} \int \frac{d^4p}{(2\pi)^4} p^x p^y \int
\frac{d^4q}{(2\pi)^4}  q^x q^y \\ \nn
&&
\times [N_p-N_{p+k}] \big(G_{aarr} + G^*_{aarr}\big)(p+k,-p,-q-k,q).
\ea
Accordingly, only $G_{aarr}$ and its complex conjugate matter when
the transport coefficients are needed.

To obtain $\eta$ and $\eta\tau_\pi$ we move forward to find $\partial_\omega \text{Im} \bar G^{xy,xy}_R$ and $\partial_\omega^2 \textrm{Re} \bar G^{xy,xy}_R$. Simultaneously, we will be applying the limit $\omega
\rightarrow 0$. So, we get
\ba
\label{der-im-tp}
&&
\lim_{\omega, k_y \rightarrow 0} 
\partial_\omega \textrm{Im} \bar G^{xy,xy}_R(\omega,k_y)
\sim \partial_\omega\Big( N_p - N_{p+k} \Big)
\Big( G_{aarr} - G^*_{aarr}\Big)(p,-p,-q,q), \\ [2mm]
\label{sec-der-4Gp}
&&
\lim_{\omega, k_y \rightarrow 0} 
\partial_\omega^2 \textrm{Re} \bar G^{xy,xy}_R(\omega,k_y)
\sim \partial^2_\omega\Big( N_p - N_{p+k} \Big)
\Big( G_{aarr} + G^*_{aarr}\Big)(p,-p,-q,q) \qquad \\ \nn
&& \qquad\qquad\qquad
+2 \partial_\omega\Big( N_p - N_{p+k} \Big)
\partial_\omega\Big( G_{aarr} + G^*_{aarr}\Big)(p+k,-p,-q-k,q).
\ea
The first line in~(\ref{sec-der-4Gp}), however, gives vanishing
contribution to the calculation of $\eta\tau_\pi$ since it is an odd function of $p_0$. 
It is seen when the following arguments are taken into account:
$n(-p_0)=-n(p_0)-1$ and $\Delta_{ar}(-p_0)=\Delta_{ra}(p_0)$.
Including the relation (\ref{N}), Eqs.~(\ref{der-im-tp}) and (\ref{sec-der-4Gp}) are rewritten as
\ba
\label{der-im-t}
&&
\lim_{\omega, k_y \rightarrow 0} 
\partial_\omega \textrm{Im} \bar G^{xy,xy}_R(\omega,k_y)
\sim 4\beta n(p_0)(n(p_0)+1) \text{Im}G_{aarr}(p,-p,-q,q), \\ [2mm]
\label{sec-der-4GF}
&&
\lim_{\omega, k_y \rightarrow 0} 
\partial_\omega^2 \textrm{Re} \bar G^{xy,xy}_R(\omega,k_y)
\sim 8\beta n(p_0)(n(p_0)+1)
\partial_\omega \text{Re}G_{aarr}(p+k,-p,-q-k,q)|_{\omega,k_y\to 0},\qquad
\ea
where $\text{Im}G_{aarr}$ and $\partial_\omega \text{Re}G_{aarr}$ are to be found 
from the Bethe-Salpeter equations for $G_{aarr}$ and $G^*_{aarr}$.
First, using the BSEs given by
(\ref{BSE-2}) and (\ref{BSE-22}), $\text{Im}G_{aarr}$ 
and $\text{Re}G_{aarr}$ are given by
\ba
\label{im-bse-p}
2 \textrm{Im} G_{pq}(0)= B_p(0) \Big[\delta_{pq}+
\int_l \mathcal{K}_{pl}(0)\, \textrm{Im} G_{lq}(0)\Big] ,
\ea
\ba
\label{re-bse}
2 \textrm{Re} G_{pq}(k)=  i A_p(k) \Big[\delta_{pq}+
\int_l \mathcal{K}_{pl}(0)\, \textrm{Im} G_{lq}(k)\Big] 
+ B_p(k) \int_l \mathcal{K}_{pl}(0)\, \textrm{Re} G_{lq}(k),
\ea
where, for clarity, we have introduced the following symbolic notations
\ba
\int_l \dots &\equiv& \int \frac{d^4l}{(2\pi)^4} \dots \\
\delta_{pq} &\equiv& (2\pi)^4 \delta^4(p-q)
\ea
and
\ba
G_{pq}(k) &\equiv& G_{aarr}(p+k,-p,-q-k,q),\\
\label{f-A}
A_p(k) &\equiv& \Delta_{ar}(p)\, \Delta_{ra}(p+k)-\Delta_{ar}(p+k)\,
\Delta_{ra}(p),\\
\label{f-B}
B_p(k) &\equiv& - \Delta_{ar}(p)\,
\Delta_{ra}(p+k)-\Delta_{ar}(p+k)\,\Delta_{ra}(p),\\
\mathcal{K}_{pl}(0) &\equiv& \mathcal{K}(p,-p,-l,l) \equiv (K_{rraa}
-N_{l}K_{rrra} +N_l K_{rrar})(p,-p,-l,l).
\ea
The imaginary part $\text{Im}G_{aarr}$ has been expressed in the vanishing $k$ limit. 
It is also the case for the kernel $\mathcal{K}_{pl}$ of the real part of $G_{aarr}$. 
This is justified due to the
following reason. In the next step we need to apply the
derivative with respect to frequency. There would appear terms
consisting of $\partial_\omega \mathcal{K}_{pl}(k)$ which are,
however, of the order of $1/\Gamma_p$ or less and they give much
smaller contribution to the final formula than the remaining ones, which are 
of the order $1/\Gamma_p^2$. Therefore, the hydrodynamic limits could have been
applied at this stage, which has simplified a lot the notation of
formula (\ref{re-bse}). The action of the derivative on Eq.~(\ref{re-bse}) in 
the hydrodynamic limits produces
\ba
\label{re-bse-der}
2 \lim_{\omega \to 0} \lim_{k_y \to 0}
\partial_\omega \textrm{Re} G_{pq}(k) &=&
 i A'_p(0) \Big[\delta_{pq} + \int_l \mathcal{K}_{pl}(0)\, \textrm{Im} G_{lq}(0)\Big] \\ \nn
&&
+i A_p(0) \int_l \mathcal{K}_{pl}(0)\, 
[\partial_\omega \textrm{Im} G_{lq}(\omega)]\big|_{\omega \to 0} \\ \nn
&&
+ B'_p(0) \int_l \mathcal{K}_{pl}(0)\, \textrm{Re} G_{lq}(0)\\ \nn
&&
+ B_p(0) \int_l \mathcal{K}_{pl}(0)\, [\partial_\omega \textrm{Re}
G_{lq}(\omega)] \big|_{\omega \to 0}
\ea
To assess which terms contribute further let us check the small
frequency behavior of $A_p$ and $B_p$ functions given by (\ref{f-A})
and (\ref{f-B}) and their derivatives,
\ba
\label{A}
A_p (0) &\equiv&  0, \\
\label{B}
B_p (0) &\equiv& - 2\Delta_{ar}(p)\, \Delta_{ra}(p), \\
\label{Ap}
A'_p(0) &\equiv& \partial_\omega A_p (\omega)\big|_{\omega \to 0} =
4i\Gamma_p(p^2_0+\Gamma^2_p+E_p^2) \Delta^2_{ar}(p)\Delta^2_{ra}(p),
\\
B'_p(0)&\equiv&\partial_\omega B_p(\omega) \big|_{\omega \to 0} =
4p_0(p^2_0+\Gamma^2_p-E_p^2) \Delta^2_{ar}(p)\Delta^2_{ra}(p).
\ea
Due to Eq. (\ref{A}),
the second term in Eq.~(\ref{re-bse-der}) vanishes. To see the
behavior of the third term we need to include the integrals as given
by the formula (\ref{GR-4-re-gen}). Then the following expression needs
to be considered:
\ba
\label{int-re}
I_3 =
\int_p I_p n_p(n_p+1)B'_p(0) F_p (0)
\ea
where $I_p = I(p)=p_x p_y$ and $F_p(0)$ is
\ba
\label{F}
F_p(0)= \int_q I_q \int_l \mathcal{K}_{pl}(0)\, \textrm{Re} G_{lq}(0).
\ea
The frequency integral part of $I_3$ is
\ba
\label{int-re-p0}
f_{3}=
\int \frac{dp_0}{2\pi} n_p(n_p+1) B'_p(0) F_p (0)
\ea
and
\ba
B'_p(0)=\frac{4p_0(p^2_0+\Gamma^2_p-E_p^2)}
{[(p_0+i\Gamma_p)^2-E_p^2]^2[(p_0-i\Gamma_p)^2-E_p^2]^2}.
\ea
The function $B'_p(0)$ has four poles at $p_1= i\Gamma_p +E_p$, 
$p_2= i\Gamma_p -E_p$, $p_3=- i\Gamma_p +E_p$, and 
$p_4=- i\Gamma_p -E_p$ and they are of the
second order. We calculate them by using the residue
theorem and closing the contour in the upper-half plane. We recall
that the residua of a function with second-order poles contain the
derivative with respect to the complex argument. 
Thus, upon carrying out the contour integration and
summing up the residua, the expression (\ref{int-re-p0}) becomes
\ba
\label{res-BF}
f_3 &= &i \lim_{p_0\to p_1} \partial_{p_0} [n_p(n_p+1)
(p_0-p_1)B'_p(0)F_p(0)] \\ \nn
&&+i\lim_{p_0\to p_2} \partial_{p_0} [n_p(n_p+1)
(p_0-p_2)B'_p(0)F_p(0)].
\ea
If we now group the terms in Eq. (\ref{res-BF}) according to the derivative with respect to
$p_0$, we can write
\ba
\label{res-BF-1}
f_3 & =&i\partial_{p_0} [n_p(n_p+1) (p_0-p_1)B'_p(0)]\big|_{p_0=p_1} 
F_{p}(0)\big|_{p_0=p_1} \nn \\
&&+ i\partial_{p_0} [n_p(n_p+1) (p_0-p_2)B'_p(0)] \big|_{p_0=p_2} 
F_{p}(0)\big|_{p_0=p_2} \nn\\
&&+i [n_p(n_p+1) (p_0-p_1)B'_p(0)]\big|_{p_0=p_1}
[\partial_{p_0}F_p(0)]\big|_{p_0=p_1} \nn\\
&&+i [n_p(n_p+1) (p_0-p_2)B'_p(0)]\big|_{p_0=p_2}
[\partial_{p_0}F_p(0)]\big|_{p_0=p_2}.
\ea
The expressions in the first and the second lines of Eq. (\ref{res-BF-1}) are
equal to 0. The other terms produce
\ba
\label{res-BF-2}
f_3 = 
\frac{n_p(n_p+1)}{E_p^2 \Gamma_p} \Big[
(\partial_{p_0}F_p(0))\big|_{p_0=p_1} +
(\partial_{p_0}F_p(0))\big|_{p_0=p_2} \Big]
\ea
which is of the order of $1/\Gamma_p$.
The derivative
$\partial_{p_0}$ in (\ref{res-BF-2}) acts only on the kernel $K_{pl}$
included in $F_p(0)$ as shown by (\ref{F}). The kernel, however, does
not contribute any terms of the order of $1/\Gamma$
so neither can do the action of $\partial_{p_0}$. The contribution of
$F_p(0)$ is $\mathcal{K}_{pl}(0)\, \textrm{Re} G_{lq}(0) \sim
\mathcal{O}(1)$. Consequently, 
the third term in Eq. (\ref{re-bse-der}) 
is only $O(1/\Gamma_p)$ and may be ignored
as we know from the one-loop analysis that $\eta\tau_\pi = O(1/\Gamma_p^2)$.

In this way we are left with the first and the fourth term of Eq.
(\ref{re-bse-der}), that is
\ba
\label{re-bse-der-ii}
\partial_\omega \textrm{Re} G_{pq}(k)\big|_{\omega,k_y \to 0}
&=&\frac{iA'_p(0)}{2}  \Big[\delta_{pq} +
\int_l \mathcal{K}_{pl}(0)\, \textrm{Im} G_{lq}(0)\Big] \\ \nn
&&
+ \frac{B_p(0)}{2} \int_l \mathcal{K}_{pl}(0)\, [\partial_\omega
\textrm{Re} G_{lq}(\omega)] \big|_{\omega,k_y \to 0}
\ea
From now on the vanishing
$\omega$ and $k_y$ limits are implicit in all expressions.

\begin{figure}[!h]
\centering
\includegraphics*[width=0.6\textwidth]{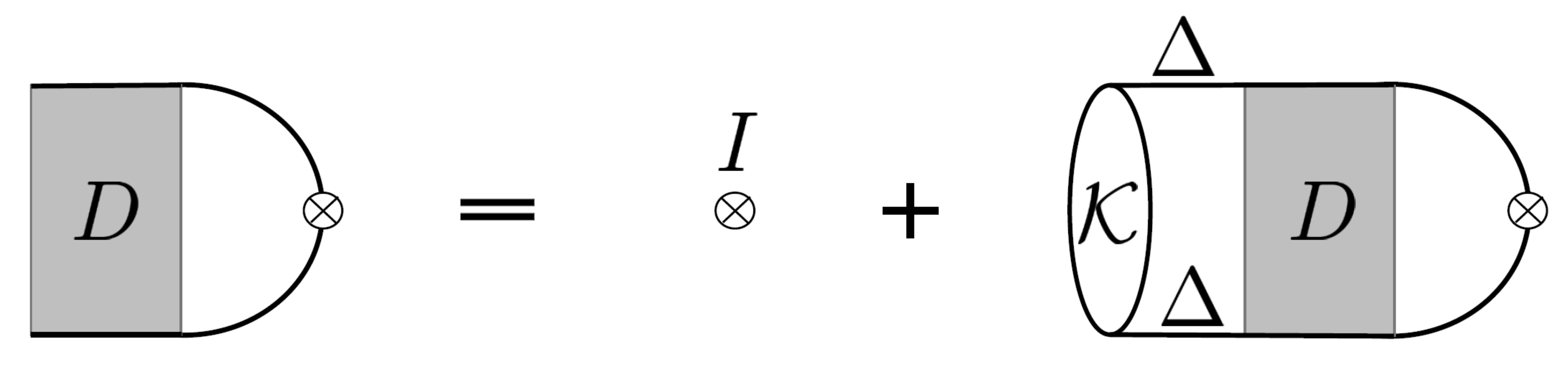}
\caption{Integral equation for the effective vertex. }
\label{fig-vertex}
\end{figure}

To reduce the BSE to a more manageable form, it is convenient to define
the effective vertex
\ba
\label{D-ver}
D_p \equiv \frac{2}{B_p}\int_q  \textrm{Im} G_{pq}\, I_q
\ea
where again
$I_q=I(q)=q_x q_y$. The effective vertex $D_p$ satisfies the following integral equation 
\ba
\label{int-eq-D}
D_p = I_p+ \int_l {\cal K}_{pl}\frac{B_l}{2}D_l
\ea
which is schematically shown in Fig.~\ref{fig-vertex}.

Then formula for $\eta$ becomes
\ba
\label{s-d}
\eta &=&
-\beta \int \frac{d^4p}{(2\pi)^4}  I_p 
n(p_0)(n(p_0)+1) B_p D_p.
\ea

For the real part, including the integral over $q$ to the equation (\ref{re-bse-der-ii})
and the expression for $D_p$, given by (\ref{int-eq-D}) we can define a new effective vertex $R_p$,
\ba
\label{eff-ver-re}
R_p \equiv \frac{2}{B_p}\int_q I_q \partial_\omega \textrm{Re} G_{pq},
\ea
which satisfies the following integral equation
\ba
\label{eff-v-H}
R_p = \frac{iA'_p}{B_p}  D_p + \int_l \mathcal{K}_{pl}\, \frac{B_l}{2} R_l
\ea
and the formula for $\eta\tau_\pi$ can be written in a compact form as
\ba
\label{compact}
\eta\tau_\pi &=& 
-\beta \int \frac{d^4p}{(2\pi)^4} I_p n(p_0)(n(p_0)+1) B_p R_p.
\ea
Note that $R_p = O(1/\Gamma_p)$.

If we insert $D_p$, given by (\ref{int-eq-D}), and $B_p$, given by 
Eq. (\ref{B}), into  Eq.~(\ref{s-d}) and perform the contour integration, 
then the shear viscosity is given by
\ba
\label{GR-4-im-gen}
\eta &=& 
\beta \int \frac{d^3p}{(2\pi)^3}  I({\bf p}) 
n(E_p)(n(E_p)+1) \frac{D(E_p,{\bf p})}{2E_p^2\Gamma_p},
\ea
where the effective vertex satisfies
\ba
\label{D-final}
D(E_p,{\bf p}) = I({\bf p}) -  \int \frac{d^3l}{(2\pi)^3}
\big(\mathcal{K}(E_p,E_l)+\mathcal{K}(E_p,-E_l)\big)\frac{D(E_l,{\bf l})}{8E_l^2 \Gamma_l}.
\ea
When deriving this we
have used $D(E_l,{\bf l})=D(-E_l,{\bf l})$, 
$\mathcal{K}(E_p,E_l)=\mathcal{K}(-E_p,-E_l)$, and 
$\mathcal{K}(-E_p,E_l)=\mathcal{K}(E_p,-E_l)$. The relations for the kernel are 
shown in Appendix \ref{notes}.

Then, using $R_p$, given by (\ref{eff-v-H}), and $B_p$ and $A'_p$, given by (\ref{B}) and (\ref{Ap}), respectively, to Eq.~(\ref{compact}), we get
\ba
\label{eta-tau-RR1}
\eta\tau_\pi = \beta \int \frac{d^3p}{(2\pi)^3}  I({\bf p}) 
n(E_p)(n(E_p)+1) \frac{R(E_p,{\bf p})}{2 E_p^2 \Gamma_p}
\ea
with the effective vertex $R(E_p,{\bf p})$ satisfying
\ba
\label{R-int-eq}
R(E_p,{\bf p}) = \frac{D(E_p,{\bf p})}{2\Gamma_p} - \int \frac{d^3l}{(2\pi)^3}
\big(\mathcal{K}(E_p,E_l)+\mathcal{K}(E_p,-E_l)\big)\frac{R(E_l,{\bf l})}{8E_l^2 \Gamma_l}.
\ea

\section{Shear relaxation time}
\label{sec-discussion}

First of all, by comparing the one-loop results
(\ref{eta-tau-fullBS-final}) and (\ref{eta-fullBS-final}), 
one clearly sees that both the shear viscosity and
the shear relaxation time are
controlled by the thermal width, $\Gamma_p$. 
The thermal width is defined by the imaginary part of the self-energy
and therefore is momentum dependent. 

Let us first, however, consider a simple example by assuming that the thermal
width is constant. Then by comparing the formulas (\ref{eta-tau-fullBS-final}) and
(\ref{eta-fullBS-final}) in the one-loop limit, it is found as
\ba
\label{tau-final-sm}
\left.\tau_\pi\right|_{\rm 1-loop} = \frac{1}{2\Gamma}.
\ea
Thus one can claim that the
thermal width $\Gamma$, as directly related to the lifetime or the mean free
path of a thermal excitation, introduces the only time scale into
the system of interacting particles and the shear relaxation time is directly related to that scale.
Hence the ratio of the one-loop results
is
\ba
\label{eta-over-tau}
\left.\frac{\eta}{\tau_\pi} \right|_{\rm 1-loop}
&=&
\beta \int \frac{d^3p}{(2\pi)^3} \frac{p_x^2
p_y^2}{E_p^2}n(E_p)(n(E_p)+1).
\ea
With the Bose-Einstein momentum distribution
$n(E_p)=1/(e^{\beta E_p}-1)$, we get
the value
\ba
\label{eta-tau-ratio-shear}
\frac{\eta}{\tau_\pi}&=& \frac{4}{450} \pi^2 T^4.
\ea
With the definition of the energy density for the one-component field,
\ba
\label{en-density-kinetic}
\langle \epsilon \rangle = 
\int \frac{d^3p}{E_p(2\pi)^3} E_p^2 \; n(E_p)
\ea
where $n(E_p)$ is the Bose-Einstein statistics and the pressure is $\langle P \rangle=\frac{1}{3} \langle \epsilon \rangle$,
the formula (\ref{eta-tau-ratio-shear}) may be rewritten in the form
\ba
\label{14-mom}
\frac{\eta}{\tau_\pi}  =\frac{\langle\epsilon+P\rangle}{5}.
\ea
The similar relation was found within 14-moment approximation to the Boltzmann equation for the classical massless gas, studied in Ref. \cite{Denicol:2014vaa}, but the energy density and thermodynamic pressure are defined there through the Boltzmann statistics. If we use the Boltzmann statistics in Eqs. (\ref{eta-over-tau}) and (\ref{en-density-kinetic}) then the relation 
(\ref{14-mom}) holds approximately.

In general, one needs to maintain the momentum dependence of the thermal width and
solve the integral equations numerically. By analyzing the integral equations for 
the effective vertices 
$D(E_p,{\bf p})$ and $R(E_p,{\bf p})$ given by Eqs.~(\ref{D-final}) and (\ref{R-int-eq}), one sees that they are of the same type but the inhomogeneous terms are different.
This indicates that the order of these two equations is different since 
$D(E_p,{\bf p})\sim \mathcal{O}(1)$ and $R(E_p,{\bf p})\sim \mathcal{O}(1/\Gamma_p)$.
Also, to find the solution to $R(E_p,{\bf p})$ one needs to first obtain the solution 
to $D(E_p,{\bf p})$. The complication of solving double integral equations can be, 
however, avoided. Define the inner product of two functions as
\be
f*g = \int {d^3p\over (2\pi)^3 2E_p^2\Gamma_p} n(E_p)(1+n(E_p)) f({\bf p})g({\bf p}).
\ee
Then the viscosity is
\be
\eta = \beta I * D.
\ee
The integral equation for $D_p$ can be symbolically written as
\be
D = I - {\cal C} * D,
\ee
where
\be
{\cal C} =
\left({\cal K}(E_p,E_l) + {\cal K}(E_p, -E_l)\right)
{1\over 4n(E_l)(1+n(E_l))}
\ee
and for $R_p$, 
\be
R = {D\over 2\Gamma} - {\cal C} * R.
\ee
Formally, the solutions are
\ba
D &=& \left(1 + {\cal C}\right)^{-1} * I,
\\
R &=& \left(1 + {\cal C}\right)^{-1} * {D\over 2\Gamma}.
\ea
The product $\eta\tau_\pi$ is then
\ba
\eta\tau_\pi &=&
\beta I * R
\nonumber
\\
&=&
\beta I *\left(1 + {\cal C}\right)^{-1} * {1\over 2\Gamma} D
\nonumber
\\
& = &
\beta D * {1\over 2\Gamma} D
\ea
provided that the kernel operator ${\cal C}$ is real and symmetric.
The same formula was found in \cite{York:2008rr} through the effective
kinetic theory approach.
The fact that ${\cal C}$
is real and symmetric is shown in Appendix \ref{notes}.

The form of the effective vertices actually reflects the fact that when 
the ladder diagrams are summed
over to get $\eta\tau_\pi$, one out of all pairs of propagators 
in each diagram contributes one 
more factor of $1/2\Gamma_p$ when compared to equivalent resummation of diagrams 
corresponding to $\eta$ calculation. Since every loop may be cut so as to represent
an elastic scattering process, one can say that each of these processes can contribute 
the lifetime associated with the momentum of incoming or outgoing particles, or also that 
of mediating particles, when loops with at least two rungs are considered. In the end
the shear relaxation time is obtained when all distinguishable possibilities are included
and they all give rise to a balanced relaxation process.

\begin{figure}[!h]
\centering
\includegraphics*[width=0.7\textwidth]{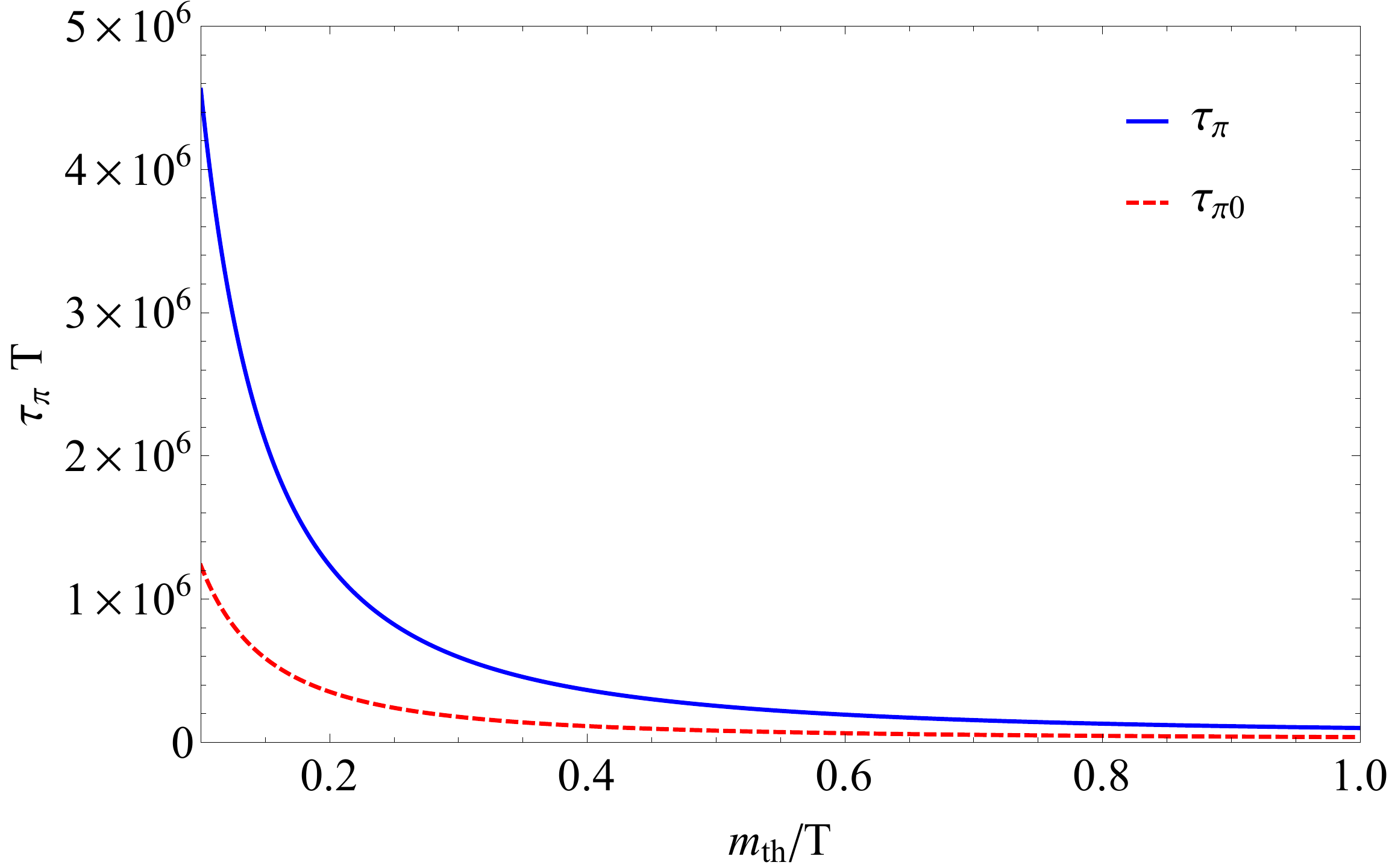}
\caption{Shear relaxation time as a function of $m_{\textrm {th}}/T$.
One-loop result ($\tau_{\pi 0}$, red dashed curve) and multi-loop resummation 
($\tau_\pi$, blue solid curve)
are presented.}
\label{fig-tau}
\end{figure}

The shear relaxation time has been then obtained by evaluating 
the integrals in (\ref{GR-4-im-gen}) and
(\ref{eta-tau-RR1}) numerically. The result is
shown in Fig.~\ref{fig-tau}, where the relaxation time is displayed as
a function of $m_{\textrm{th}}/T$ with $m_{\textrm{th}}$ being the
thermal mass, which introduces the natural cut off to infrared
divergences. The ratio $m_{\textrm{th}}/T$ ought to be identified with
 $\sqrt{\lambda}$ since the thermal mass behaves as $m^2_{\textrm{th}}=\lambda T^2$ in 
the leading order and the range of the plot has been chosen
in such a way to show a general behavior of the shear relaxation time. Yet, one needs to
keep in mind that when the coupling constant is increasing, nonperturbative
effects start to play more and more a role and the value of the relaxation time
becomes less and less realistic. 
On the other hand, our analysis equally well
applies even if the physical mass is non-zero.
Hence, the large $m_{\rm th}/T$ part of the results shown in this section
can be interpreted as those for the the massive scalar field 
but still weakly coupled. 

The solid (blue) curve in Fig.~\ref{fig-tau} corresponds to the shear relaxation time
obtained as the ratio of (\ref{eta-tau-RR1}) over
(\ref{GR-4-im-gen}). The dashed (red) curve represents the shear
relaxation time evaluated directly from one-loop expressions (\ref{eta-tau-fullBS-final}) and
(\ref{eta-fullBS-final}).  
As can be seen, the difference between
these two treatments is noticeable. One also
immediately observes that the bigger the coupling constant is, the
shorter the shear relaxation time becomes, as expected. What is more,
the shear relaxation time is around three times bigger than the corresponding one-loop finding
so it is not justified to claim that the one-loop result 
dominates the behavior of the relaxation time.

\begin{figure}[!h]
\centering
\includegraphics*[width=0.7\textwidth]{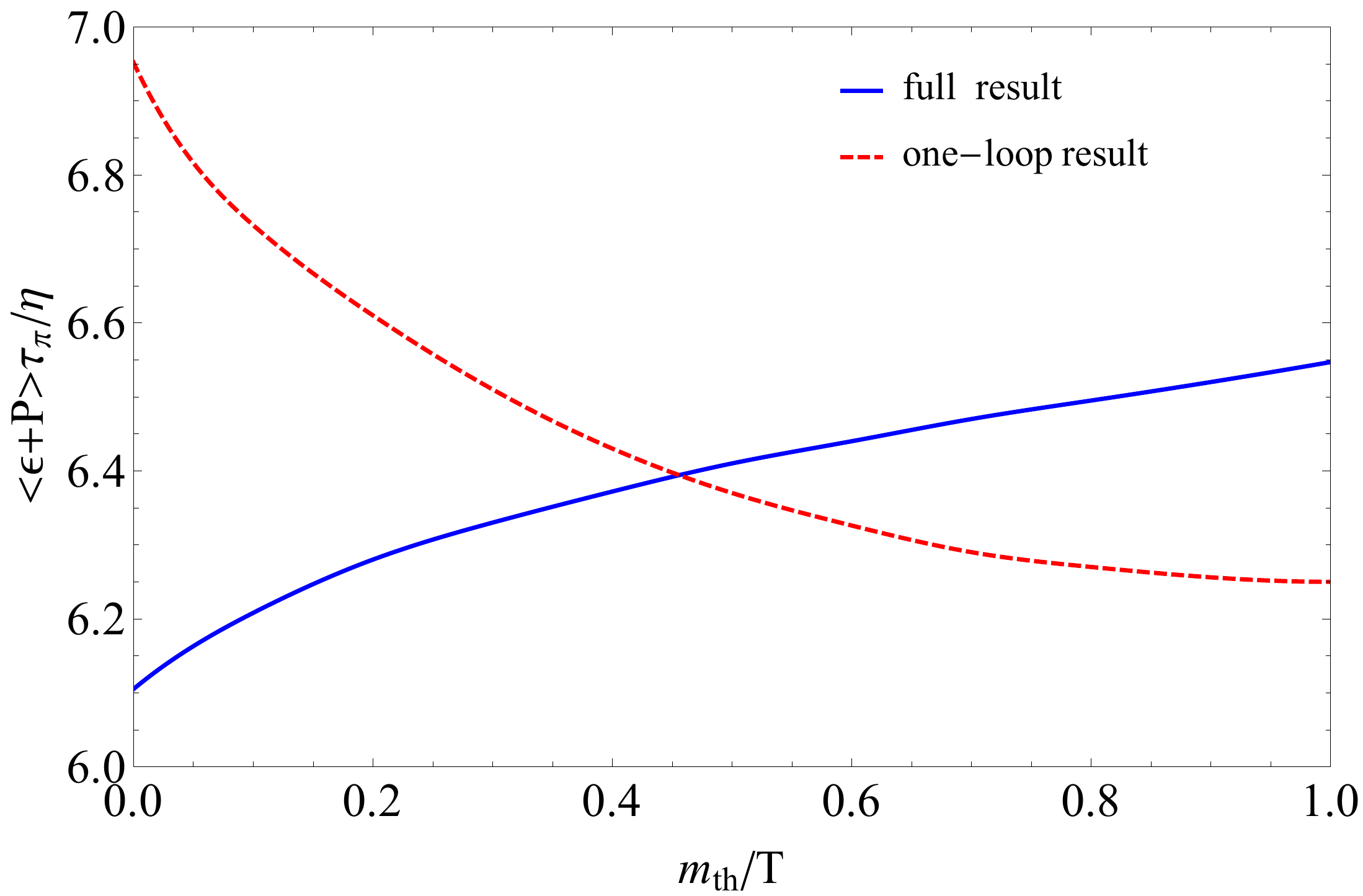}
\caption{The ratio $\langle\epsilon+P\rangle\tau_\pi/\eta$ as a function of $m_{\textrm
{th}}/T$. Evaluation in the one-loop limit (red dashed curve) and multi-loop
resummation (blue solid curve) are presented. }
\label{fig-tau-shear}
\end{figure}

We also present the ratio $\langle\epsilon+P\rangle\tau_\pi/\eta$ as a function of $m_{\textrm{th}}/T$, which is shown in
Fig.~\ref{fig-tau-shear}. One can see that the ratio is decreasing when
coupling constant decreases and it varies between 6.11 up to 6.55 in the range shown. Furthermore, the tendency of the one-loop result is just opposite.

The ratio $\langle\epsilon+P\rangle\tau_\pi/\eta$ as a function of $m_{\textrm{th}}/T$
was also obtained in \cite{York:2008rr} within the effective kinetic theory 
for QCD, QED, and scalar $\lambda \phi^4$ theory. In the case of the scalar theory
only the value at $m_{\textrm{th}} = 0$ was presented in \cite{York:2008rr}.
As $m_{\textrm{th}}/T\to 0$, they observed
this ratio to be 6.11 for the full theory
and exactly 6 for the massless Boltzmann gas. 
As can be seen, the quantum field theoretical findings obtained here are in line with these 
of the effective kinetic theory
in the regime of a very small coupling constant and one may expect that this equivalence
 holds in the whole range of $m_{\textrm{th}}/T$.

The ratio $\langle\epsilon+P\rangle\tau_\pi/\eta$ obtained here for the scalar theory behaves in opposite way than the one obtained in \cite{York:2008rr} for QCD in the range of 
$m_{\textrm{th}}/T$ changing from 0 to 1. When the coupling constant increases the nonperturbative effects become, as mentioned, more and more essential but even for 
very small values of the coupling different nature of the interaction governing both
theories should be taken into consideration. In QCD Coulomb-like interaction dominates the 
physics of the system which results in appearance of $\textrm{ln}(1/g)$, with $g$ being the coupling 
constant of strong interaction, in the parametric form of the
transport coefficients. What is more, the cross section is strongly angle dependent which 
gives rise to the soft and near collinear singularities. As argued in \cite{York:2008rr} the
collinear splittings become more and more important when the coupling constant increases and
they make the ratio $\langle\epsilon+P\rangle\tau_\pi/\eta$ go down. In the scalar
$\lambda \phi^4$ theory the contact interaction causes the cross-section to be 
isotropic therefore only
ladder diagrams constitute the leading order behavior of the shear viscosity and its relaxation time.
In Ref. \cite{Jeon:1994if} it was shown that indeed the soft and near collinear singularities 
are of smaller size than the ladder diagrams and they have not had to be discussed in detail here.
A broader analysis of the next-to-leading order behavior of shear viscosity is given
in Ref.~\cite{Moore:2007ib}, where the soft physics (quasiparticle momenta $k$ being 
of the order $m_{\textrm{th}}$) is shown to determine the NLO correction. Since the leading order equation does not differ in the form from the one that includes subleading corrections, we expect
that higher order contributions will behave similarly, but we leave that for future analysis.

It is also illuminating to notice that the leading order behavior of different transport 
coefficients manifests different susceptibility to the thermal mass and to the way it emerges in the computational analysis. For instance, the thermal mass plays an essential
role in the case of bulk viscosity where not only the leading order term but also higher order 
corrections in the coupling constant must be properly incorporated. 
Then, apart from the loops with an arbitrary number 
of rungs inside them, there appear so-called chain diagrams made of an arbitrary number of loops.
Addition of a subsequent loop introduces an additional coupling constant. However, although 
chain diagrams emerge at every possible order in the coupling constant their resummation
is of the size $\mathcal{O}(\lambda T^2)$, that is, the same as the thermal mass squared. 
Therefore, the chain diagrams give a significant contribution to the bulk effects analysis. In the case
of the shear viscosity or shear relaxation time calculation this effect is absent since rotational 
symmetry makes the chain diagrams vanish, as discussed in \cite{Jeon:1994if}. Accordingly, these higher order corrections do not have to be considered here.

\section{Conclusions}
\label{sec-conclusions}

The first goal in this paper was to work out Kubo
formulas for the shear and the bulk relaxation times.
Since the Kubo formula for the shear relaxation time was studied before, our 
focus was on the Kubo formula which allows us to compute the bulk relaxation time. 
Our Kubo formula is different than the one obtained in \cite{Huang:2010sa} 
using the projection operator method, but it is consistent with the one found
in \cite{Hong:2010at} using the metric perturbation method.

Although our ultimate goal is to compute the bulk relaxation time,
in this paper we have concentrated on simpler task of computing
the shear relaxation time. This is primarily to check the soundness of
our overall formulations and refine the field theoretical techniques we will
need for the much more involved bulk relaxation time computation.
Calculation of the bulk relaxation time is in progress.

The behavior of the shear relaxation time obtained in this study is
consistent with previous studies within the kinetic theory setting
as well as within the field theory setting. 
Our conclusion that
\be
{\tau_\pi \langle \epsilon + P\rangle\over \eta}
\approx 5 - 7
\ee
seems to be robust across different theories and also consistent with
the values used in hydrodynamic calculations \cite{Vujanovic:2016anq}.

\section*{Acknowledgments}

A. Czajka acknowledges support from the program Mobility Plus of the
Polish Ministry of Science and Higher Education.
S.J. was supported in part by the Natural Sciences and
Engineering Research Council of Canada.
Discussions with C.~Gale and G.D.~Moore are very much appreciated.

\appendix

\section{Ward identity analysis}
\label{app:ward_id}

The stress-energy tensor Ward identity 
(\ref{Ward-id-real}) is
\ba
k_\alpha \big( \bar G^{\alpha\beta,\mu\nu}(k) 
- g^{\beta\mu} \langle \hat T^{\alpha\nu} \rangle 
- g^{\beta\nu} \langle \hat T^{\alpha\mu} \rangle 
+ g^{\alpha\beta} \langle \hat T^{\mu\nu} \rangle \big).
=0
\ea
In the local rest frame of the medium, the above expression can be decomposed as
\ba
&& 
\omega(\bar{G}^{00,00}(k) - \langle\epsilon\rangle)
=
- k_i\bar{G}^{0i,00} (k),
\label{eq:f1}
\\
&& 
\omega \big( \bar G^{00,0i }(k) \big)
=
-
k_j \big( \bar G^{0j, 0i }(k) - \delta_{ij} P \big),
\label{eq:f2}
\\
&& 
\omega \big( \bar G^{00,{ij}}(k) + \delta_{ij}P  \big)
=
-
k_l \big( \bar G^{0l,{ij}}(k) \big),
\label{eq:f3}
\\
&& 
\omega \big( \bar G^{0j,00 }(k) \big)
=
-
k_i \big( \bar G^{{ij},00 }(k) - \delta_{ij} \epsilon \big),
\label{eq:f4}
\\
&& 
\omega \big( \bar G^{0j,0i }(k) + \delta_{ij} \epsilon \big)
=
-
k_l \big( \bar G^{{lj},0i }(k) \big),
\label{eq:f5}
\\
&& 
\omega \big( \bar G^{0j,{lm} }(k) \big)
=
-
k_i \big( \bar G^{{ij},{lm} }(k) 
+ \delta^{j l } \delta^{im} P 
+ \delta^{j m } \delta^{il} P
- \delta^{i j } \delta^{l m} P \big).
\label{eq:f6}
\ea
In the ${\bf k}\to 0$ limit, the right hand sides 
of Eqs. (\ref{eq:f1})--(\ref{eq:f6}) must vanish. Hence,
\ba
&& \bar{G}^{\epsilon\epsilon}(\omega,0) = \epsilon,
\\
&& \bar{G}^{\epsilon s^{ij}}(\omega, 0) = -\delta_{ij}P,
\\
&&
\bar G^{\pi^j\pi^i}(\omega, 0) = -\delta_{ij}\epsilon,
\ea
In the $\omega\to 0$ limit, the left hand side must vanish.
Hence, 
\ba
&&\bar{G}^{\pi^j \pi^i}(0,{\bf k}) = \delta_{ij}P 
+ \hat{\delta}_{ij}g_{\pi\pi}({\bf k}),
\label{eq:app_pi_pi}
\\
&&\bar{G}^{s^{ij}\epsilon}(0,{\bf k}) = \delta_{ij}\epsilon
+ \hat{\delta}_{ij}g_{s\epsilon}({\bf k}),
\label{eq:app_s_eps}
\\
&&
\bar{G}^{s^{ij}s^{lm}}(0,{\bf k})
= 
- \big (\delta^{j l } \delta^{im}  + \delta^{j m }
- \delta^{i j } \delta^{l m} \big) P 
+ g_{ijlm}({\bf k}),
\label{eq:app_s_s}
\ea
where $\hat{\delta}_{ij} = \delta_{ij} - \hat{k}_i\hat{k}_j$
with $\hat{k}_i = k_i/|{\bf k}|$ is the transverse projection
and $g_{ijlm}$ is transverse with respect to all its indices.
Each of the $g_{\cdots}({\bf k})$ functions must be at least
$O({\bf k}^2)$ in the small ${\bf k}$ limit so that $\hat{k}_i\hat{k}_j$ in
$\hat{\delta}_{ij}$ is well defined.

We also need $\bar{G}^{\epsilon\epsilon}(0,{\bf k})$.
In the zero frequency limit, the retarded correlation function
and the Euclidean one coincide. Hence,
\ba
\lim_{{\bf k}\to 0}G_E^{\epsilon\epsilon}(0,{\bf k})
& = &
\int d^3x\int_0^\beta d\tau\,
\langle \hat{T}^{00}(\tau,{\bf x})\hat{T}^{00}(0)\rangle_{\rm conn} +
\epsilon
\nonumber\\
& = &
\beta
\left( \langle \hat{H} T^{00} \rangle -
\langle\hat{H}\rangle\langle T^{00}\rangle \right) + \epsilon.
\ea
Note that
\be
{\partial\over \partial T} e^{-\beta\hat{H}}
= \beta^2 \hat{H} e^{-\beta\hat{H}},
\ee
which leads to
\be
\lim_{{\bf k}\to 0}G_E^{\epsilon\epsilon}(0,{\bf k})
= 
T {\partial \epsilon\over \partial T} + \epsilon,
\label{eq:app_c_v}
\ee
where $\partial\epsilon/\partial T = c_v$ is the specific
heat per unit volume.

\section{Decomposition of stress-stress correlation function}
\label{app:decomp}

Consider $\bar{G}^{ij,lm}$ which is the correlation function of 
$\hat{T}^{ij}$ and $\hat{T}^{lm}$. When the space is isotropic, there are following five 
independent tensors that respect the symmetry
\begin{eqnarray}
&&\delta_{ij}\delta_{lm},\ \ 
\delta_{il}\delta_{jm} +\delta_{im}\delta_{jl},\ \ 
\delta_{ij}\hat{k}_l\hat{k}_m + \delta_{lm}\hat{k}_i\hat{k}_j, 
\nonumber\\
&&
\delta_{il}\hat{k}_j\hat{k}_m
+
\delta_{jm}\hat{k}_i\hat{k}_l
+
\delta_{im}\hat{k}_j\hat{k}_l
+
\delta_{jl}\hat{k}_i\hat{k}_m,\ \ 
\hat{k}_i\hat{k}_j\hat{k}_l\hat{k}_m,
\end{eqnarray}
which are composed of $\delta_{ij}$ and $\hat{k}_i = k_i/|\bf{k}|$.
Hence, there are altogether five independent functions that can appear
in $\bar{G}^{ij,lm}$. Since we are interested in the shear and the bulk responses,
it is more convenient to define the transverse metric
\begin{equation}
\hat{\delta}_{ij} = \delta_{ij} - \hat{k}_i\hat{k}_j.
\end{equation}
The stress-stress correlation function can be then decomposed
as \cite{LGY}
\begin{eqnarray}
\bar{G}^{ij,lm}(\omega,{\bf k})
& = &
-P\left(\delta_{il}\delta_{jm} + \delta_{im}\delta_{jl} 
- \delta_{ij}\delta_{jm}\right)
\nonumber\\
& & {}
+ \bar G_1(\omega,{\bf k})
\left(
\hat\delta_{il}\hat{k}_j\hat{k}_m
+
\hat\delta_{jm}\hat{k}_i\hat{k}_l
+
\hat\delta_{im}\hat{k}_j\hat{k}_l
+
\hat\delta_{jl}\hat{k}_i\hat{k}_m 
\right)
\nonumber\\
& & {}
+ \bar G_2(\omega,{\bf k})
\left(
\hat\delta_{il}\hat\delta_{jm} +\hat\delta_{im}\hat\delta_{jl}-
\hat\delta_{ij}\hat\delta_{lm}
\right)
\nonumber\\
& & {}
+ \bar G_T(\omega,{\bf k})\, \hat\delta_{ij}\hat\delta_{lm}
\nonumber\\
& & {}
+
\bar G_{LT}(\omega,{\bf k})
\left(
\hat\delta_{ij}\hat{k}_l\hat{k}_m +
\hat\delta_{lm}\hat{k}_i\hat{k}_j\right)
\nonumber\\
& & {}
+ \bar G_L(\omega,{\bf k})\, \hat{k}_i\hat{k}_j\hat{k}_l\hat{k}_m.
\label{eq:Gijlm}
\end{eqnarray}

This expression must be well defined in the ${\bf k} \to 0$ limit.
Hence, the coefficient of, say, $\delta_{ij}\hat{k}_l\hat{k}_m$
must be $O({\bf k}^2)$ as ${\bf k}\to 0$.
Collecting the coefficients, one sees that
\begin{equation}
\bar G_2 - \bar G_T + \bar G_{LT} = O({\bf k}^2).
\end{equation}
The coefficient of $\delta_{il}\hat{k}_j\hat{k}_m$ is
\begin{equation}
- \bar G_2 + \bar G_1 = O({\bf k}^2).
\end{equation}
The coefficient of $\hat{k}_i\hat{k}_j \hat{k}_l\hat{k}_m$ must be
$O({\bf k}^4)$,
\be
2 \bar G_2 - 4 \bar G_1 + \bar G_T + \bar G_L - 2 \bar G_{LT} = O({\bf k}^4).
\ee
Hence in the ${\bf k} \to 0$ limit, only two functions are independent. 
Choosing those to be $\bar G_1$ and $\bar G_L$, we get
\begin{eqnarray}
&& \bar G_2(\omega,0) = \bar G_1(\omega,0),
\nonumber\\
&& \bar G_T(\omega,0) = \bar G_L(\omega,0) - \bar G_1(\omega,0),
\nonumber\\
&& \bar G_{LT}(\omega,0) = \bar G_L(\omega,0) - 2 \bar G_1(\omega,0).
\end{eqnarray}

From Eq. (\ref{eq:Gijlm}), one can easily see that 
$\bar{G}_R^{xy,xy}(\omega, k_y) = -P + \bar G_1(\omega, k_y)$
and
$\bar{G}_R^{xy,xy}(\omega, k_z) = -P + \bar G_2(\omega, k_z)$.
From Ref. \cite{Baier:2007ix}, we know that in the small $k_z$ limit,
\begin{equation}
\bar G_2(\omega, k_z) = -i\eta\omega + \tau_\pi\eta\omega^2 -
{\kappa\over 2}(\omega^2 + k_z^2) + \hbox{higher orders}.
\end{equation}
However, since $\bar G_1(\omega, {\bf k}) =\bar G_2(\omega, {\bf k}) + O({\bf k}^2)$,
one cannot in general say that $\bar{G}_R^{xy,xy}(\omega,k_z)$ behaves
the same as $\bar{G}_R^{xy,xy}(\omega, k_y)$ nor the same as
$\bar{G}_R^{xy,xy}(\omega, k_x)$.

To get the Kubo formulas for the bulk viscosity and the bulk
relaxation time, the pressure-pressure correlation function is
needed. In the limit ${\bf k} \to 0$, Eq. (\ref{eq:Gijlm}) becomes,
with $\hat{P} = \delta_{ij}\hat{T}^{ij}/3$,
\begin{eqnarray}
\bar{G}_R^{PP}(\omega, 0) 
& = &
{P\over 3}
+ {4\over 9}\bar G_T(\omega, 0) 
+ {4\over 9}\bar G_{LT}(\omega,0)
+ {1\over 9}\bar G_L(\omega,0)
\nonumber\\
& = &
{P\over 3}
+\bar G_L(\omega,0) - {4\over 3}\bar G_1(\omega,0)
\nonumber\\
& = &
{P\over 3}
+ i\omega (\zeta + 4\eta/3) 
- (\zeta\tau_\Pi + 4\eta\tau_\pi/3 - 2\kappa/3)\omega^2
-{4\over 3}(i\omega\eta -\eta\tau_\pi\omega^2 + \kappa\omega^2/2)
+ O(\omega^3)
\nonumber\\
& = &
{P\over 3} + i\omega\zeta - \zeta\tau_\Pi\omega^2
+ O(\omega^3).
\end{eqnarray}

\section{Closed time path formalism}
\label{KS-form}

Here we briefly describe the closed time path or Keldysh-Schwinger
formalism, which is studied in more detail, for example, in
\cite{Chou:1984es}. The main object of the formalism is the contour
Green function which has four components of the real-time arguments. Here,
we define them for the scalar field operators $\phi$ but these
definitions may be directly generalized to any composite field operators, such as
$\hat T^{ij}$ discussed in the main body of the paper. The components
of the contour Green function are given as
\ba
\label{12}
\Delta_{a_1a_2}(x,y) &=& -i \langle \mathcal{T} \phi_{a_1}(x)
\phi_{a_2}(y)\rangle,
\ea
where $a_1,a_2 \in \{1,2\}$ and the indices 1 and 2 refer to the two
branches of the Keldysh contour the field operator $\phi$ is located
on. The operator $\mathcal{T}$ represents an ordering of the operators
along the contour; $\mathcal{T}=\mathcal{T}_c$ chronologically orders the
operators on the upper branch and $\mathcal{T}=\mathcal{T}_a$ sets
anti-chronological ordering on the lower branch. For the angle brackets we
use the following notation:
\ba 
\label{trace}
\langle \dots \rangle \equiv \frac{{\textrm {Tr}} [\hat\rho(t_0)
\dots]}{{\textrm {Tr}} [\hat\rho(t_0)]}
\ea
with $\hat\rho(t_0)$ being a density operator and the trace is
understood as a summation over all states of the system at a given initial time
$t_0$. The averaged products of unordered operators are commonly known as
the Wightman functions. All the functions in the (1,2) basis satisfy
the
relation
\ba
\label{rel-12}
\Delta_{11}+\Delta_{22}=\Delta_{12}+\Delta_{21},
\ea
which reflects the fact that only three out of four components are
independent of each other.

Going to the $(r,a)$ basis, we define
\ba
\label{r-a-def}
\phi_a(x) = \phi_1(x) - \phi_2(x) ,\qquad\qquad\qquad
\phi_r(x)=\frac{1}{2}\Big[\phi_1(x)+\phi_2(x)\Big]
\ea
and then the four components are defined by
\ba
\label{12}
\Delta_{\alpha_1\alpha_2}(x,y) &=& -i 2^{n_r-1}\langle \mathcal{T}
\phi_{\alpha_1}(x) \phi_{\alpha_2}(y)\rangle,
\ea
where $\alpha_1,\alpha_2 \in \{r,a\}$ and $n_r$ is a number of $r$
indices among $\alpha_1$ and $\alpha_2$.

For further purposes it is useful to know the relations between the
Green functions of $(r,a)$ and $(1,2)$ bases, which read
\ba
\label{rr-12}
\Delta_{rr}(x,y) &=& \Delta_{12}(x,y) + \Delta_{21}(x,y),\\
\label{ra-12}
\Delta_{ra}(x,y) &=& \Delta_{11}(x,y) - \Delta_{12}(x,y), \\ \nn
&=& \theta(x_0-y_0) [\Delta_{12}(x,y) - \Delta_{21}(x,y)],\\
\label{ar-12}
\Delta_{ar}(x,y) &=& \Delta_{11}(x,y) - \Delta_{21}(x,y), \\ \nn
&=& -\theta(y_0-x_0) [\Delta_{12}(x,y) - \Delta_{21}(x,y)],\\
\label{aa-12}
\Delta_{aa}(x,y) &=& 0,
\ea
where $\theta(x_0)$ is the step function. A general transformation
law which holds for any $n$-point Green function is then
\ba
\label{tr-law-n-point}
G_{\alpha_1 \dots \alpha_n}(x_1, \dots, x_n)= 2^{\frac{n}{2}-1}G_{a_1
\dots
a_n}(x_1, \dots, x_n) Q_{\alpha_1 a_1} \dots Q_{\alpha_n a_n},
\ea
where the repeated indices are summed over and
$Q_{a1}=-Q_{a2}=Q_{r1}=Q_{r2}=\frac{1}{\sqrt{2}}$ are the four
elements of the orthogonal Keldysh transformation.

The inverted relations (\ref{rr-12})--(\ref{aa-12}) read
\ba
\label{11-ra}
\Delta_{11}(x,y) &=& \frac{1}{2} \big(\Delta_{rr}(x,y) +
\Delta_{ra}(x,y) +
\Delta_{ar}(x,y) \big),\\
\Delta_{12}(x,y) &=& \frac{1}{2} \big(\Delta_{rr}(x,y) -
\Delta_{ra}(x,y) +
\Delta_{ar}(x,y) \big),\\
\Delta_{21}(x,y) &=& \frac{1}{2} \big(\Delta_{rr}(x,y) +
\Delta_{ra}(x,y)
- \Delta_{ar}(x,y) \big), \\
\label{22-ra}
\Delta_{22}(x,y) &=& \frac{1}{2} \big(\Delta_{rr}(x,y) -
\Delta_{ra}(x,y)
- \Delta_{ar}(x,y) \big) 
\ea
and the general transformation from $(1,2)$ to $(r,a)$ basis for any
$n$-point function is then
\ba
\label{invert-rel}
G_{a_1 \dots a_n}(x_1, \dots, x_n)= 2^{1-\frac{n}{2}} G_{\alpha_1
\dots
\alpha_n}(x_1, \dots, x_n) Q_{\alpha_1 a_1} \dots Q_{\alpha_n a_n}.
\ea

The functions $\Delta_{ra}(x,y)$ and $\Delta_{ar}(x,y)$ are the usual
retarded and advanced Green functions. In case of massless theory they are
of the following forms in the momentum space
\ba
\label{ret-adv-fun}
\Delta_{ra}(k)&=&\frac{1}{(k_0+i\Gamma_k)^2 - E_k^2},\\
\Delta_{ar}(k)&=&\frac{1}{(k_0-i\Gamma_k)^2 - E_k^2},
\ea
and they satisfy
\ba
\label{ret-adv-rell-1}
\Delta_{ra}(k)=\Delta_{ar}^*(k).
\ea
$\Delta_{rr}(k)$ is the correlation function and it is the only
function where a distribution function enters. In thermal equilibrium all
these three functions are related via the fluctuation-dissipation
theorem
\ba
\label{fl-dis-theorem-2}
\Delta_{rr}(k)=[1+2n(k^0)][\Delta_{ra}(k)-\Delta_{ar}(k)],
\ea
where $n(k^0)=1/(e^{\beta k^0}-1)$ is the Bose distribution function
with
$\beta$ being the inverse of temperature $T$.

By means of the relation~(\ref{ret-adv-rell-1}), we immediately find
real and imaginary parts of the retarded propagator
\ba
\label{real-part-ra}
2\, \textrm{Re}\;\Delta_{ra} (k) = \Delta_{ra}(k) + \Delta_{ar}(k),\\
\label{imaginary-part-ra}
2i\, \textrm{Im}\;\Delta_{ra} (k) = \Delta_{ra}(k) - \Delta_{ar}(k).
\ea

\section{Four-point Green functions}
\label{notes}

Here we provide some useful derivations arising during the analysis of the
Bethe-Salpeter equation.

\begin{center}
{\bf Relations among different four-point Green functions}
\end{center}

The four-point Green functions which contribute to the real and imaginary parts of
$\bar G_R^{xy,xy}$ are $G_{arrr}$, $G_{rarr}$, $G_{rrar}$,
and $G_{rrra}$. The analysis of the Bethe-Salpeter equation gets
easier when one realizes that these functions may be expressed by
$G_{aarr}$ and $G_{rraa}$. Let us first notice that by amputating two
external legs out of the $G_{aarr}$ function from the left hand side
one may
write
\ba
\label{Gaarr}
G_{aarr}(p+k,-p,-q-k,q) 
= 
i\Delta_{ar}(p+k) i\Delta_{ar}(-p) M_{rrrr}(p+k,-p,-q-k,q),
\ea
and by amputating two external legs of $G_{rraa}$ from the right-hand
side one gets
\ba
\label{Grraa}
G_{rraa}(p+k,-p,-q-k,q) 
= 
i\Delta_{ra}(q+k) i\Delta_{ra}(-q) M_{rrrr}(p+k,-p,-q-k,q).
\ea
Using the expressions (\ref{Gaarr}) and (\ref{Grraa}), we can express
$G_{rrra}$, $G_{rrar}$, $G_{rarr}$, and $G_{arrr}$ as follows 
\ba
\label{Garrr}
G_{arrr}(p+k,-p,-q-k,q) 
&=& 
i\Delta_{ar}(p+k) i\Delta_{rr}(-p) M_{rrrr}(p+k,-p,-q-k,q)\\ \nn
&=&
N_p G_{aarr}(p+k,-p,-q-k,q), \\
\label{Grarr}
G_{rarr}(p+k,-p,-q-k,q) 
&=&
i\Delta_{rr}(p+k) i\Delta_{ar}(-p) M_{rrrr}(p+k,-p,-q-k,q)\\ \nn 
&=&
-N_{p+k}G_{aarr}(p+k,-p,-q-k,q), \\
G_{rrar}(p+k,-p,-q-k,q) 
&=& 
i\Delta_{ra}(q+k) i\Delta_{rr}(-q) M_{rrrr}(p+k,-p,-q-k,q)\\ \nn 
&=&
-N_q G_{rraa}(p+k,-p,-q-k,q), \\
\label{Grrra}
G_{rrra}(p+k,-p,-q-k,q) 
&=& 
i\Delta_{rr}(q+k) i\Delta_{ra}(-q) M_{rrrr}(p+k,-p,-q-k,q)\\ \nn
&=&
N_{q+k}G_{rraa}(p+k,-p,-q-k,q), 
\ea
where $N_p=N(p_0)$ and we also have used the identity
$N(-p_0)=-N(p_0)$. Then, we see
\ba
G_{arrr}+G_{rarr}&=& [N_p-N_{p+k}]G_{aarr}, \\
G_{rrar}+G_{rrra}&=& [-N_q+N_{q+k}]G_{rraa}.
\ea

\begin{center}
{\bf Kernel of the Bethe-Salpeter equation}
\end{center}

The general expression of any rung contributing to the kernel is 
\ba
\label{kernel-gen-ex}
K_{\beta_1 \gamma_1\beta_4\gamma_4}(p+k,-p,-l-k,l) 
= 
\frac{1}{2} \int \frac{d^4s}{(2\pi)^4}
\lambda_{\gamma_1 \gamma_2 \gamma_3 \gamma_4} 
\lambda_{\beta_1 \beta_2 \beta_3 \beta_4}
\Delta_{\gamma_2 \beta_2}(s) 
\Delta_{\gamma_3 \beta_3}(s-l+p)
\ea
and the bare 4-point vertex is given by
\ba
\label{vertex}
\lambda_{\gamma_1 \gamma_2 \gamma_3 \gamma_4}
=
\frac{\lambda}{4}[1-(-1)^{n_a}],
\ea
where $n_a$ is the number of $a$ indices among $\gamma_1 \gamma_2
\gamma_3 \gamma_4$. With the definitions (\ref{kernel-gen-ex}) and
(\ref{vertex}), the three rungs under interest are given by
\ba
\label{Krrra}
&&K_{rrra}(p+k,-p,-l-k,l) \\ \nn
&& \qquad
=\frac{\lambda^2}{8} \int \frac{d^4s}{(2\pi)^4}
\Big[\Delta_{rr}(s) \Delta_{ar}(s-l+p)
+\Delta_{ra}(s)\Delta_{rr}(s-l+p)
\Big], \\
\label{Krrar}
&&K_{rrar}(p+k,-p,-l-k,l) \\ \nn
&& \qquad
=\frac{\lambda^2}{8} \int \frac{d^4s}{(2\pi)^4}
\Big[\Delta_{ar}(s) \Delta_{rr}(s-l+p)
+\Delta_{rr}(s)\Delta_{ra}(s-l+p)
\Big], \\
\label{Krraa}
&&K_{rraa}(p+k,-p,-l-k,l) \\ \nn
&& \qquad
=\frac{\lambda^2}{8} \int \frac{d^4s}{(2\pi)^4}
\Big[\Delta_{rr}(s) \Delta_{rr}(s-l+p)
+\Delta_{ra}(s)\Delta_{ar}(s-l+p)
+ \Delta_{ar}(s)\Delta_{ra}(s-l+p) \Big].
\ea
Then, the following relations hold
\ba
\label{KK}
K^*_{rraa}&=&K_{rraa}, \\ 
K^*_{rrra}&=&-K_{rrar}, \\
K^*_{rrar}&=&-K_{rrra}
\ea
which are helpful in investigation of the Bethe-Salpeter equation for
$G_{aarr}^*$. In the vanishing frequency and momentum limits, the
kernel may be denoted as
\ba
\label{kernel-den1}
\mathcal{K} = K_{rraa} +N_{l}K_{rrar} -N_l K_{rrra}.
\ea

The kernel, as explained in \cite{Wang:2002nba}, may be also expressed in a more
convenient form as
\ba
\label{kernel-rho}
\mathcal{K}(p,-p,-l,l) = -\frac{\lambda^2}{2}
\frac{1+n(l_0)}{1+n(p_0)}
\int \frac{d^4s}{(2\pi)^4} n(s_0) \rho(s) [1+n(s_0-l_0+p_0)]
\rho(s-l+p),
\ea
where $\rho(s)=i[\Delta_{ra}(s) - \Delta_{ar}(s)]$. If we denote 
$\mathcal{K}(p,-p,-l,l)=\mathcal{K}(p,l)$, the following relations for the kernel hold:
\ba
\label{ker-po}
\mathcal{K}(-p,-l)
& =& \mathcal{K}(p,l)
\ea
and
\ba
\label{ker-pl}
\mathcal{K}(-p,l)
&=& \mathcal{K}(p,-l)
\ea
both of which can be shown by changing the integration variable from $s$ to $-s$.
One can also show
\ba
\mathcal{K}(p,l) 
&=&{n(l_0)\over n(p_0)}{1 + n(l_0)\over 1 + n(p_0)}{\cal K}(l,p)
\ea
by changing the integration variable $s$ to $s-l+p$.


\end{document}